overlap%

\documentclass[twoside]{emulateapj}
\pdfoutput=1

\usepackage{graphicx}
\usepackage{rotating}
\usepackage{url}
\usepackage{natbib}
\usepackage{xcolor}
\usepackage{appendix}

\newcommand{\Lsun}{{\rm\thinspace L_\odot}}

\slugcomment{}

\shorttitle{SERVS: survey definition and goals}
\shortauthors{Mauduit et al.}

\begin{document}


\title{The Spitzer Extragalactic Representative Volume Survey (SERVS): survey
  definition and goals}


\author{
  J.-C.\ Mauduit\altaffilmark{1}, M.\ Lacy\altaffilmark{2},
  D.\ Farrah\altaffilmark{3}, J.A.\ Surace\altaffilmark{1},
  M.\ Jarvis\altaffilmark{4}, S.\ Oliver\altaffilmark{3}, 
  C.\ Maraston\altaffilmark{5},
  M.\ Vaccari\altaffilmark{6,7}, L.\ Marchetti\altaffilmark{6},
  G.\ Zeimann\altaffilmark{8},  E.A.~Gonz\'{a}les-Solares\altaffilmark{9}, 
  J.\ Pforr\altaffilmark{5,10}, A.O.\ Petric\altaffilmark{1},
  B.\ Henriques\altaffilmark{2}, P.A.\ Thomas\altaffilmark{2},
  J.\ Afonso\altaffilmark{11,12}, 
  A.\ Rettura\altaffilmark{13}, G.\ Wilson\altaffilmark{13},     
  J.T.\ Falder\altaffilmark{4}, J.E.\ Geach\altaffilmark{14},
  M.\ Huynh\altaffilmark{15}, R.P.\ Norris\altaffilmark{16}, 
  N.\ Seymour\altaffilmark{16}, G.T.\ Richards\altaffilmark{17}, 
  S.A.\ Stanford\altaffilmark{8,18}, 
  D.M.\ Alexander\altaffilmark{19}, R.H.\ Becker\altaffilmark{8,18},
  P.N.\ Best\altaffilmark{20}, L.\ Bizzocchi\altaffilmark{11,12},
  D.\ Bonfield\altaffilmark{4}, 
  N.\ Castro\altaffilmark{21}, A.\ Cava\altaffilmark{21},
  S.\ Chapman\altaffilmark{9}, N.\ Christopher\altaffilmark{22},
  D.L.\ Clements\altaffilmark{23}, G. Covone,\altaffilmark{24},
  N.\ Dubois\altaffilmark{3},
  J.S.\ Dunlop\altaffilmark{20}, E.\ Dyke\altaffilmark{4},
  A.\ Edge\altaffilmark{25}, 
  H.C.\ Ferguson\altaffilmark{26}, S.\ Foucaud\altaffilmark{27},
  A.\ Franceschini\altaffilmark{6}, R.R.\ Gal\altaffilmark{28},
  J.K.\ Grant\altaffilmark{29},
  M.\ Grossi\altaffilmark{11,12}, E.\ Hatziminaoglou\altaffilmark{30},
   S.\ Hickey\altaffilmark{4},
  J.A.\ Hodge\altaffilmark{31}, J.-S.\ Huang\altaffilmark{31},
  R.J.\ Ivison\altaffilmark{20}, M.\ Kim\altaffilmark{1},
  O.\ LeFevre\altaffilmark{32}, M.\ Lehnert\altaffilmark{33},
  C.J.\ Lonsdale\altaffilmark{1}, L.M.\ Lubin\altaffilmark{8},
  R.J.\ McLure\altaffilmark{20}, H.\ Messias\altaffilmark{11,12},
  A.\ Mart\'{i}nez-Sansigre\altaffilmark{5,22},
  A.M.J.\ Mortier\altaffilmark{20}, D.M.\ Nielsen\altaffilmark{34},
  M.\ Ouchi\altaffilmark{35},
  G.\ Parish\altaffilmark{4}, I.\ Perez-Fournon\altaffilmark{21},
  M.\ Pierre\altaffilmark{36}, S.\ Rawlings\altaffilmark{22},
  A.\ Readhead\altaffilmark{37}, 
  S.E.\ Ridgway\altaffilmark{38},
  D.\ Rigopoulou\altaffilmark{22},
  A.K.\ Romer\altaffilmark{2}, I.G.\ Roseboom\altaffilmark{2},
  H.J.A.\ Rottgering\altaffilmark{39}, M.\ Rowan-Robinson\altaffilmark{23},
  A.\ Sajina\altaffilmark{40}, 
  C.J.\ Simpson\altaffilmark{41}, I.\ Smail\altaffilmark{25},
  G.K.\ Squires\altaffilmark{1}, J.A.\ Stevens\altaffilmark{4},
  R.\ Taylor\altaffilmark{29}, 
  M.\ Trichas\altaffilmark{23}, T.\ Urrutia\altaffilmark{42}, E.\ van
  Kampen\altaffilmark{29}, A.\ Verma\altaffilmark{22},
  C.K.\ Xu\altaffilmark{1}
}

\altaffiltext{1}{Infrared Processing and Analysis Center/Spitzer Science
  Center, California Institute of Technology, Mail Code 220-6, Pasadena, CA
  91125, USA} 
\altaffiltext{2}{National Radio Astronomy Observatory, 520 Edgemont Road,
  Charlottesville, VA 22903, USA}  
\altaffiltext{3}{Department of Physics and Astronomy, University of Sussex,
  Falmer, Brighton, BN1 9QH, UK}
\altaffiltext{4}{Center for Astrophysics Research, University of
  Hertfordshire, Hatfield, AL10 9AB, UK}
\altaffiltext{5}{Institute of Cosmology and Gravitation, University of
  Portsmouth, Dennis Sciama Building, Burnaby Road, Portsmouth, PO1 3FX, UK}  
\altaffiltext{6}{Department of Astronomy, Universit\`{a} di Padova, Vicolo dell'Osservatorio 3, 35122,
  Padova, Italy}
\altaffiltext{7}{Astrophysics Group, Physics Department, University of the
  Western Cape, Private Bag X17, 7535, Bellville, Cape Town, South Africa} 
\altaffiltext{8}{Department of Physics, University of California, One Shields
  Avenue, Davis, CA95616, USA} 
\altaffiltext{9}{Institute of Astronomy, University of Cambridge, Madingley
  Road, Cambridge, CB3 0HA, UK}  
\altaffiltext{10}{National Optical Astronomy Observatory, 950 
North Cherry Avenue, Tuscon, AZ 85719, USA}  
\altaffiltext{11}{Observat\'{o}rio Astron\'{o}mico de Lisboa, Faculdade de
  Ci\^{e}ncias, Universidade de Lisboa, Tapada da Ajuda, 1349-018 Lisbon,
  Portugal}  
\altaffiltext{12}{Centro de Astronomia e Astrof\'{\i}sica da Universidade de
  Lisboa, Tapada da Ajuda, 1349-018 Lisbon, Portugal} 
\altaffiltext{13}{Department of Physics and Astronomy, University of
  California-Riverside, 900 University Avenue, Riverside, CA 92521, USA} 
\altaffiltext{14}{Department of Physics, McGill University, Ernest Rutherford
  Building, 3600 rue University, Montr\'{e}al, Qu\'{e}bec H3A 2T8, Canada} 
\altaffiltext{15}{International Centre for Radio Astronomy Research,
  University of Western Australia, M468, 35 Stirling Highway, Crawley WA 6009,
  Australia} 
\altaffiltext{16}{Commonwealth Scientific and Industrial Research Organisation, Astronomy \& Space Science, PO Box 76, Epping, NSW,
  1710, Australia} 
\altaffiltext{17}{Department of Physics, Drexel University, 3141 Chesnut
  Street, Philadelphia, PA 19014, USA} 
\altaffiltext{18}{Institute of Geophysics and Planetary Physics, Lawrence Livermore National Laboratory, 7000 East
  Ave., Livermore, CA94550, USA} 
\altaffiltext{19}{Department of Physics, University of Durham, South Road,
  Durham, DH1 3LE, UK} 
\altaffiltext{20}{Institute for Astronomy, University of Edinburgh, Royal
  Observatory, Blackford Hill, Edinburgh, EH9 3HJ, UK}
\altaffiltext{21}{Institutio de Astrof\'{i}sica de Canarias, C/V\'{i}a
  L\'{a}ctea s/n, 38200, La Laguna, Tenerife, Spain}
\altaffiltext{22}{Oxford Astrophysics, Denys Wilkinson Building, Keble Road,
  Oxford, OX1 3RH, UK}  
\altaffiltext{23}{Astrophysics Group, Blackett Laboratory, Imperial College, Prince
Consort Road, London, SW7 2BW, UK}
\altaffiltext{24}{Dipartimento di Scienze Fisiche, Universit\`a Federico II
  and Istituto Nazionale di Fisica Nucleare, Sezione di Napoli, Complesso
  Universitario di Monte S. Angelo, Via Cintia, Edificio 6, I-80126 Napoli, Italy} 
\altaffiltext{25}{Institute for Computational Cosmology, Durham University,
  South Road, Durham, DH1 3LE, UK} 
\altaffiltext{26}{Space Telescope Science Institute, 3700 San Martin Drive,
  Baltimore, MD 21218, USA} 
\altaffiltext{27}{Department of Earth Sciences, National Taiwan Normal University,  N$^{\circ}$88, Tingzhou Road, Sec. 4, Taipei 11677, Taiwan (R.O.C.)} 
\altaffiltext{28}{Institute for Astronomy, University of Hawaii, 2680
  Woodlawn Drive, Honolulu, HI 96822, USA} 
\altaffiltext{29}{Institute for Space Imaging Science, University of Calgary,
  AB T2N 1N4, Canada} 
\altaffiltext{30}{European Southern Observatory, Karl-Schwartzschild-Strasse 2, 85748, Garching,
Germany}
\altaffiltext{31}{Max-Planck Institute for Astronomy, Konigstuhl 17, 69177,
  Heidelberg, Germany} 
\altaffiltext{32}{Laboratoire d'Astrophysique de Marseille, Traverse du
  Siphon, B.P.8, 13376 Marseille Cedex 12, France}
\altaffiltext{33}{Laboratoire d'Etudes des Galaxies, Etoiles, Physique et
  Instrumentation GEPI, UMR8111, Observatoire de Paris, Meudon, 92195,
  France} 
\altaffiltext{34}{Astronomy Department, University of Wisconsin, Madison, 475
  North Charter Street, Madison, WI 53711, USA}
\altaffiltext{35}{Observatories of the Carnegie Institute of Washington, 813
  Santa Barbara Street, Pasadena, CA 91101, USA}
\altaffiltext{36}{Commissariat \`{a} l'Energie Atomique, Saclay, F-91191 Gif-sur-Yvette, France}
\altaffiltext{37}{Astronomy Department, California Institute of Technology,
  Mail Code 247-17, 1200 East California Boulevard, Pasadena, CA 91125, USA}   
\altaffiltext{38}{Cerro Tololo Interamerican Observatory, Colina El Pino s/n,
  Casilla 603, La Serena, Chile} 
\altaffiltext{39}{Leiden Observatory, Leiden University, Oort Gebouw, PO Box
  9513, 2300 RA Leiden, The Netherlands}
\altaffiltext{40}{Department of Physics and Astronomy, Haverford College,
  Haverford, PA, 19041, USA} 
\altaffiltext{41}{Astrophysics Research Institute, Liverpool John Moores
  University, Twelve Quays House, Egerton Wharf, Birkenhead CH41 1LD}  
\altaffiltext{42}{Leibniz Institute for Astrophysics, An der Sternwarte 16, 14482 Potsdam, Germany} 

\begin{abstract}

We present the {\em Spitzer Extragalactic Representative Volume
Survey} (SERVS), an $18\,\rm{deg}^2$ medium-deep survey at 3.6 and $4.5\, \mu$m
with the post-cryogenic {\em Spitzer Space Telescope} to $\approx$\,2\,$\mu$Jy
($AB=23.1$) depth of five highly observed astronomical fields (ELAIS-N1, ELAIS-S1,
Lockman Hole, Chandra Deep Field South and XMM-LSS). SERVS is designed to enable the study of galaxy
evolution as a function of environment from $z\sim 5$ to the present day, and
is the first extragalactic survey that is both large enough and deep enough to put
rare objects such as luminous quasars and galaxy clusters at
$z\stackrel{>}{_{\sim}}1$ into their cosmological context. SERVS is designed
to overlap with several key surveys at optical, near- through far-infrared,
submillimeter and radio wavelengths to provide an unprecedented view of the
formation and evolution of massive galaxies. In this article, we discuss the SERVS survey
design, the data processing flow from image reduction and mosaicking to
catalogs, and coverage of ancillary data from other surveys in the
SERVS fields. We also highlight a variety of early science results from
the survey. 

\end{abstract}



\section{Introduction}
\label{sec:intro}

\setcounter{footnote}{0}

Progress in extragalactic astronomy has been greatly enhanced by deep surveys
such as the Great Observatories Origins Deep Survey (GOODS, 
\citealt{Dickinson+03}), the Cosmic Evolution Survey (COSMOS,
\citealt{Sanders+07}), the Galaxy Mass Assembly ultradeep Spectroscopic
Survey (GMASS, \citealt{Cimatti+08}), the {\em HST} Cosmic
Assembly Near-infrared Deep Extragalactic Legacy Survey (CANDELS, \citealt{Grogin+11,Koekemoer+11}), 
that have allowed us to study the evolution of galaxies from the
earliest cosmic epochs. However, a limitation of such surveys is the
relatively small volumes probed, even at high redshifts: for example,
\cite{Ilbert+06} find noticeable field-to-field variations in redshift
distributions in the Canada-France-Hawaii Telescope Legacy Survey
(CFHTLS\footnote{www.cfht.hawaii.edu/Science/CFHTLS}) in fields of
$0.7-0.9\,$deg$^2$.  

Until lately, the combination of depth and area required to map a large
volume ($\sim 1\,$Gpc$^3$) of the high-redshift Universe at near-infrared
wavelengths, where the redshifted emission from stars in distant galaxies
peaks, has been prohibitively expensive in telescope time. Two recent
developments have now made this regime accessible. On the ground, the
availability of wide-field near-infrared cameras has greatly improved the
effectiveness of ground-based near-infrared surveys in the $1 - 2.5\, \mu$m wavelength
range. In space, the exhaustion of the cryogenic coolant of the {\em Spitzer
  Space Telescope} opened up an opportunity to pursue large near-IR surveys using the
two shortest wavelength channels (IRAC1 [3.6] and IRAC2 [4.5]) of the Infrared Array
Camera (IRAC, \citealt{Fazio+04}) in the post-cryogenic or ``warm'' mission
that were much larger than was feasible during the cryogenic mission. The {\em
  Spitzer Extragalactic Representative Volume Survey} (SERVS), a {\em
  Spitzer} ``Exploration Science'' program, stems from these two
developments.  

\begin{figure*}
\begin{minipage}[b]{1\linewidth}
\centering
\vspace{1.5cm}
\begin{tabular}{cc}
\hspace{-1cm}\includegraphics[trim=0cm 0.7cm 0cm 0cm,clip=true,scale=0.75]{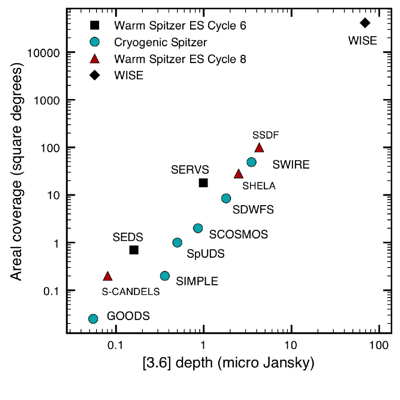} &
\hspace{0.5cm}\includegraphics[trim=0cm 0.37cm 0cm 0cm,clip=true,scale=1.115]{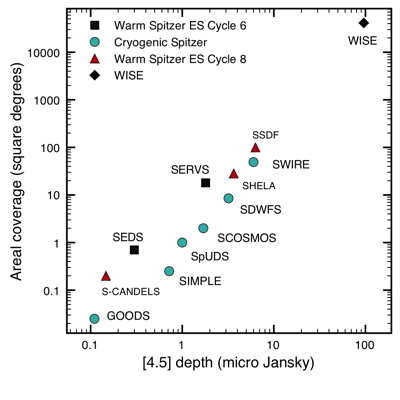}\\
\end{tabular}
\end{minipage}
\caption{Area versus depth for SERVS compared to other surveys at wavelengths
  of $\approx 3.6 \mu$m (\emph{left panel}) and $\approx 4.5 \mu$m
  (\emph{right panel}).
For consistency, the depth shown is
  the $5\,\sigma$  
  limiting flux for point sources, excluding confusion noise ($\sigma_{\rm
    pp}$ as described in Section~\ref{subsec:data_analysis}), calculated from the {\em Spitzer}
  performance estimation tool 
  (http://ssc.spitzer.caltech.edu/warmmission/propkit/pet/senspet
  /index.html) in each case. The surveys are ({\it from left to right}):
  GOODS, the {\em Spitzer} follow-up to the
CANDELS {\em HST} survey (Cosmic Assembly Near-IR Deep Extragalactic Legacy Survey,
\citealt{Grogin+11,Koekemoer+11}), the {\em Spitzer} Extragalactic Deep Survey 
  (SEDS, Program identifier - hereafter PID - 60022, 61040, 61041, 61042,
61043, P.I.\ G.\ Fazio), the {\em Spitzer} IRAC/MUSYC Public Legacy in E-CDFS
(SIMPLE) survey ({\em Spitzer}, PID 20708), the {\em Spitzer} Ultra Deep
Survey (SpUDS, PID 40021, P.I.\ J.S.\ Dunlop), S-COSMOS, the {\em Spitzer}
Deep Wide-Field Survey (SDWFS, \citealt{Ashby+09}), the {\em Spitzer}-HETDEX
Exploratory Large Area (SHELA, PID 80100, P.I.\ C.\ Papovich) Survey, SWIRE,
the SPT-{\em Spitzer} Deep Field (SSDF, PID 80096, P.I.\ S.\ Stanford) and
the Wide-Field Infrared Explorer (WISE, \citealt{Wright+10}).}
\label{fig:areadepth}
\end{figure*}

SERVS is designed to open up a medium-depth, medium-area part of parameter 
space in the near-infrared (see Figure~\ref{fig:areadepth}), covering 18\,deg$^2$
to $\approx 2\, \mu$Jy in the {\em Spitzer} [3.6] and [4.5]
bands. These observations required 1400hr of telescope time and covered five highly
observed astronomical fields: ELAIS-N1 (hereafter EN1), ELAIS-S1 (ES1), Lockman Hole (Lockman), Chandra Deep
Field South (CDFS) and XMM-large-scale structure (XMM-LSS). The five SERVS fields are centered on or close to
those of corresponding  fields surveyed by the shallower {\em Spitzer}
Wide-area Infrared Extragalactic Survey (SWIRE;  
\citealt{Lonsdale+03}), and overlap with several other major
surveys covering wavelengths from the X-ray to the radio. 
Of particular importance is near-infrared data, as these allow accurate
photometric redshifts to be obtained for high redshifts
\citep{vanDokkum+06,BvDC08,Ilbert+09,Cardamone+10}: SERVS overlaps exactly
with the 12\,deg$^2$ of the VISTA Deep Extragalactic Observations (VISTA
VIDEO, Jarvis et al.\ 2012, in preparation) survey ($Z, Y, J, H$ and $K_s$ bands) in
the south, and is covered by the UKIRT Infrared Deep Sky Survey (UKIDSS
  DXS, \citealt{Lawrence+07}) survey ($J$, $K$) in the 
north. SERVS also has good overlap with the {\em Herschel} Multitiered
Extragalactic Survey (HerMES, \citealt{Oliver+12}) in the far-infrared, which covers the SWIRE
and other {\em Spitzer} survey fields, with deeper subfields within many of
the SERVS fields. 

Sampling a volume of $\sim 0.8$\,Gpc$^3$ from redshifts 1 to 5, the survey is
large enough to contain significant numbers of rare objects, such as luminous 
quasars, ultraluminous infrared galaxies (ULIRGs), radio galaxies and galaxy
clusters, while still being deep enough to find $L^*$ galaxies out to
$z\approx 5$ (see for example \citealt{Falder+11}, and \citealt{Capak+11} who
find two galaxies in the $z=5.3$ cluster 
bright enough to be detected by SERVS at 4.5$\mu$m.)  
For comparison, the largest structures seen in the Millennium
simulation at $z\sim 1$ are of the order of $100$ Mpc \citep{Springel+05}, which
subtends $3^{\circ}$ at that redshift, so each SERVS field samples a wide range
of environments. By combining the five different fields of SERVS, the survey
effectively averages over large-scale structure, and presents a representative picture
of the average properties of galaxies in the high redshift Universe.

{\em Spitzer} observations of the five SERVS fields are presented in detail in
Section~\ref{sec:observations}. Image processing is detailed in
Section~\ref{sec:processing}, focusing on the mosaicking process and uniformity of
coverage. Section~\ref{sec:catalogs} presents the
extracted SERVS catalogs, as well as an assessment of overall data quality,
detection limits and expected number counts. 
Section~\ref{sec:ancillary} gives an
overview of the ancillary data available at different wavelengths in the five
fields. Preliminary science results and science goals are described in Section~\ref{sec:science_goals}. 
A summary of the SERVS data at hand is provided in
Section~\ref{sec:summary}.

\vspace{0.5cm}

\section{{\em Spitzer} observations}
\label{sec:observations}

\subsection{Selection of fields}
\label{subsec:fieldselection}

\begin{table*}
\centering
\caption{The geometry of the SERVS fields}
\begin{tabular}{lccccccc}
\hline
\hline
Field Name & Field Center & Field PA & Field Area & \multicolumn{4}{c}{Vertices$^*$ of the area covered by both [3.6] \& [4.5]} \\ 
  &   RA, Dec (J2000) &  (deg)  &  (deg$^2$)   & \multicolumn{4}{c}{(deg)} \\
\hline
EN1 & 16:10:00,\,$+$54:30& 350& 2.0  &(244.2,54.2) & (243.1,55.4)&(240.9,54.8)&(241.7,53.6) \\
ES1 & 00:37:48,\,$-$44:00& 0  & 3    &(10.5,-44.9) & (10.4,-42.9)&(8.4,-43.0)&(8.4,-45.1)   \\
Lockman  & 10:49:12,\,$+$58:07& 328& 4.0  &(165.0,57.4) & (161.7,59.8)  &(159.3,59.0)&(162.7,56.4)\\
CDFS     & 03:32:19,\,$-$28:06& 0  & 4.5  &(54.4,-27.1) & (51.8,-27.0)&(51.7,-28.9)&(54.4,-28.9)\\
XMM-LSS  & 02:20:00,\,$-$04:48& 0  & 4.5  &(37.2,-5.4)  & (37.0,-3.9)&(33.9,-4.1)&(34.3,-5.7)\\
\hline
\end{tabular}
\\
\scriptsize \hspace{-9cm} $^*$ Single-band catalogs extend beyond vertices. 
\label{table:servsgeometry}
\vspace{0.5cm}
\end{table*}

\begin{figure}
\resizebox{\hsize}{!}{\includegraphics[trim=0cm 0cm 0.8cm 0.7cm,clip=true]{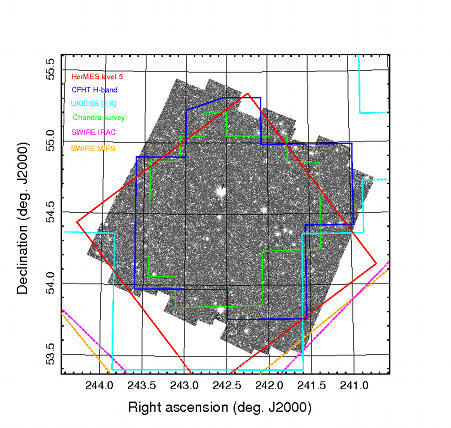}}
\caption{The [3.6] SERVS mosaic image of the EN1 field. Surveys of
  comparable size are shown here, such as HerMES level 5 (\emph{in red}, see \citealt{Oliver+12} for details), CFHT
  H-band (\emph{dark blue}), UKIDSS J \& K-bands (\emph{light blue}), the
  Chandra X-ray survey (\emph{green}), SWIRE IRAC (\emph{dashed magenta}) and
  SWIRE MIPS
  (\emph{dashed orange}). Surveys encompassing the entire SERVS field 
  such as HerMES Level 6 and the GMRT survey at 610 MHz are not shown
  here. More details about ancillary data coverage can be found in 
  Section~\ref{sec:ancillary}.}
\label{fig:en1fig}
\end{figure}

\begin{figure}
\resizebox{\hsize}{!}{\includegraphics[trim=0cm 0.5cm 0.7cm 0.2cm,clip=true]{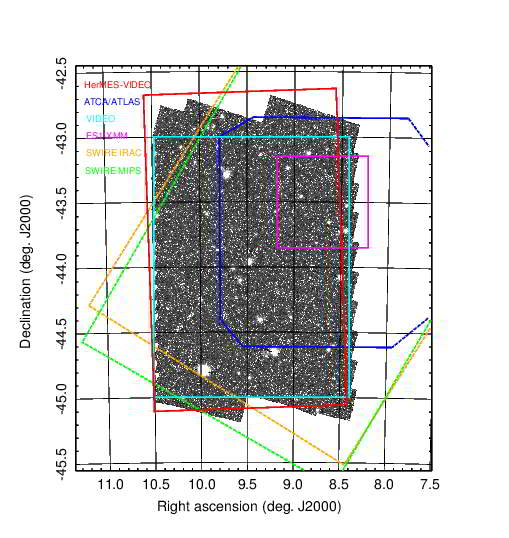}}
\caption{The [3.6] SERVS mosaic image of the ES1 field. Surveys of
    comparable size are shown here, such as the HerMES--VIDEO field (\emph{in
      red}), the ATCA/ATLAS radio survey (\emph{dark blue}), the VIDEO survey
    (\emph{light blue}), the deep ES1-XMM field of \cite{Feruglio+08}
    (\emph{magenta}), SWIRE IRAC (\emph{dashed orange}) and SWIRE MIPS
    (\emph{dashed green}). Surveys encompassing the entire field such as
    HerMES Level 6 are not shown here. More details about ancillary data
    coverage can be found in Section~\ref{sec:ancillary}.}
\label{fig:es1fig}
\end{figure}

\begin{figure}
\hspace{0.2cm}
\resizebox{\hsize}{!}{\includegraphics[trim=0cm 0cm 0.4cm 0.5cm,clip=true]{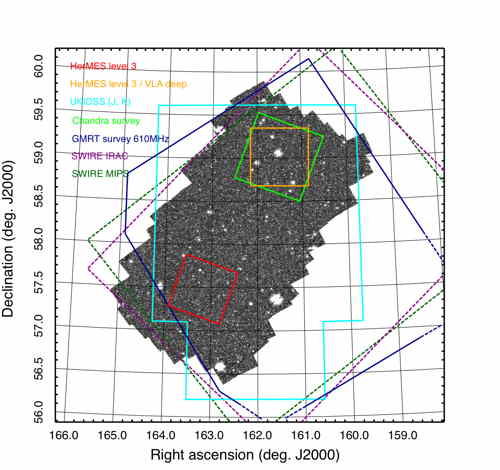}} 
\caption{The [3.6] SERVS mosaic image of the Lockman field. Only surveys of
  comparable sizes are shown here. Superposed are the HerMES Level 5 (\emph{in
    red}) and Level 3 (\emph{magenta \& orange}), the Owen/Wilkes deep VLA
  (\emph{orange}), the UKIDSS $J, K$ coverage (\emph{cyan}). The
  \emph{Chandra} survey is displayed in \emph{green} and the GMRT survey in
  \emph{blue}. SWIRE IRAC \& MIPS are shown as \emph{dashed dark magenta} and
  \emph{dashed dark green}. Surveys encompassing the entire field such as
  HerMES Level 5 are not shown here. More details about ancillary data
  coverage can be found in Section~\ref{sec:ancillary}.}
\label{fig:lockmanfig}
\vspace{0.2cm}
\end{figure}

\begin{figure}
\resizebox{\hsize}{!}{\includegraphics[trim=0cm 0cm 0.4cm 0cm,clip=true]{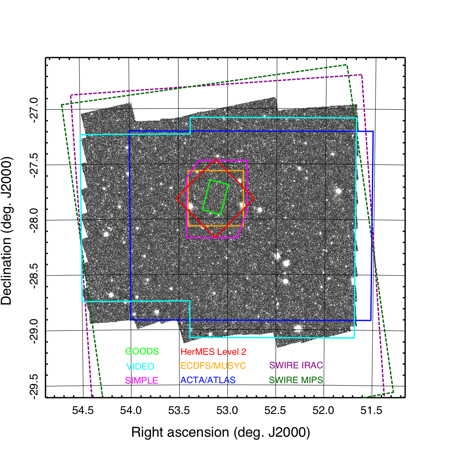}}
\caption{The [3.6] SERVS mosaic image of the CDFS field. Shown here are the
  HerMES Level 2 pointing (\emph{in red}), the ECDFS/MUSYC survey
  (\emph{orange}), the SIMPLE \& GOODS surveys (\emph{magenta} and
  \emph{green} respectively). The VIDEO pointing is shown in \emph{cyan} and
  the ATLAS radio survey in \emph{blue}. SWIRE IRAC \& MIPS are
  shown as \emph{dashed dark magenta} and \emph{dashed dark green}. Surveys
  encompassing the entire 
  field such as HerMES Level 5 are not shown here. More
  details about ancillary data coverage can be found in
  Section~\ref{sec:ancillary}.} 
\label{fig:cdfsfig}
\vspace{0.2cm}
\end{figure}

\begin{figure}
\resizebox{\hsize}{!}{\includegraphics[trim=0.2cm 0cm 1.4cm 0.5cm,clip=true]{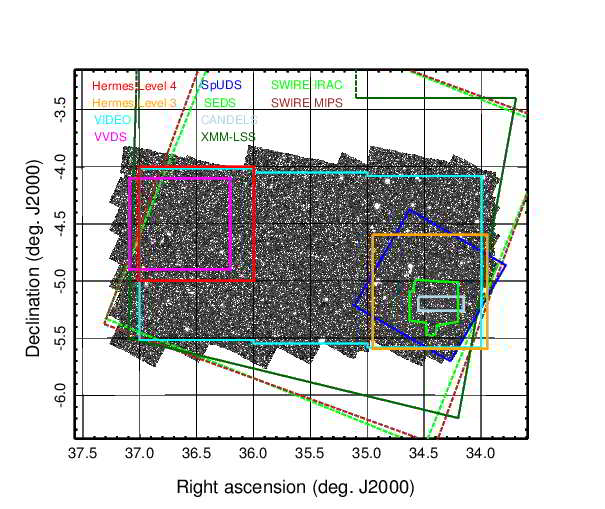}} 
\caption{The [3.6] SERVS mosaic image of the XMM-LSS field. The two HerMES
  Level 4 \& 3 pointings are shown in \emph{red} and \emph{orange}. The VIDEO
  pointing is featured in \emph{cyan} and VVDS in \emph{magenta}. The Eastern
  fields correspond to the SpuDS (\emph{blue}), SEDS (\emph{green}) and
  CANDELS (\emph{light grey}). XMM-LSS is shown in (\emph{dark green}) and
  extends beyond the SERVS CDFS field limits. SWIRE IRAC is shown as
  \emph{dashed green} and SWIRE MIPS as \emph{dashed brown}. Surveys
  encompassing the entire field such as HerMES Level 6 are not shown
  here. More details about ancillary data coverage can be found in
  Section~\ref{sec:ancillary}.}
\vspace{1cm}
\label{fig:xmmfig}
\end{figure}

SERVS consists of five fields located near the centers of corresponding SWIRE
fields: EN1, ES1, Lockman, XMM-LSS and CDFS. The SWIRE fields are in regions with 
low infrared backgrounds \citep{Lonsdale+03}, 
making them ideal for follow-up at far-infrared wavelengths. The
SERVS fields were selected to have good overlap with current and proposed
surveys in other wavebands within the SWIRE fields (see Section~\ref{sec:intro}), to cover both northern
and southern hemispheres, and to have a range in Right Ascension allowing
both flexible follow-up with ground-based telescopes and good scheduling
opportunities for {\em Spitzer}. Field geometry and observation details are
given in Table~\ref{table:servsgeometry} \&
Table~\ref{table:servsfields}. The observed SERVS mosaics are shown
in Figures~\ref{fig:en1fig} through \ref{fig:xmmfig}, together with the
coverage of significant overlapping surveys (see Section~\ref{sec:ancillary}
for an exhaustive list of all ancillary data in and near the SERVS fields).

A small fraction of the SERVS area was already covered by other deep surveys
with IRAC, such as the {\em Spitzer} IRAC/MUSYC Public Legacy in E-CDFS
survey (SIMPLE, PID 20708; PI P.\ van Dokkum; \citealt{Damen+09}) in the CDFS
field and the {\em Spitzer} Ultra Deep Survey (SpUDS, PID 40021; PI
J.S.\ Dunlop) in XMM-LSS. In order to minimize the total required telescope
time, these particular areas were not observed. A selection of the IRAC [3.6]
and [4.5] data from these surveys (both of which also use the $100\,$s
frametime) are therefore subsequently added into the final SERVS mosaics to
attain an approximately uniform overall depth (see details in
Section~\ref{subsec:imcov}). 

In addition to this pre-existing {\em Spitzer} data, there are
also two smaller deep fields located in the SERVS area: AORID\footnote{An
  individual \emph{Spitzer} observation sequence is an Astronomical Observation
  Request (AOR)} 4402688 in Lockman (PID
64; P.I.\ Fazio) and the overlapping pointings of AORIDs 6005016 (PID 196,
P.I.\ Dickinson) \& 10092288 (PID 3407, P.I.\ Yan) in
EN1. We deliberately reimaged them as part of SERVS so that the data taken during
the post-cryogenic period could be compared with the data taken earlier in the mission, and their small size made tiling around them very inefficient.

\subsection{Design of observations}
\label{subsec:AOR_design}

The design of the SERVS observations reflected several tradeoffs to ensure
efficient use of the telescope, accurate filling of the fixed field
geometries, and reasonably flexible scheduling. 
The SERVS depth was selected so that the confusion level just became
significant; attempts to make it much deeper would require better
ancillary data (e.g., GOODS) reaching in the confusion noise (the rate
at which depth is achieved no longer decreases as the square root of
exposure time. Within
the constraints of the call for proposals, at this depth, SERVS is the largest
area that can easily be surveyed and that has matching ancillary data. The
depth of SERVS allows us to detect all massive ($>10^{11} M_{\odot}$)  
galaxies out to $z\sim 4$ (see section 4.3), essentially the entire range of
redshift over which they are seen. SERVS can thus trace the evolution of 
these objects from their formation until the present epoch. As a consequence
of those factors, the survey covers 18\,deg$^2$ and reaches down to $\approx
2\,\mu$Jy in the {\em Spitzer} [3.6] and [4.5] bands. 

\begin{figure}
\begin{minipage}[b]{1\linewidth}
\centering
\vspace{0.2cm}
\includegraphics[scale=1]{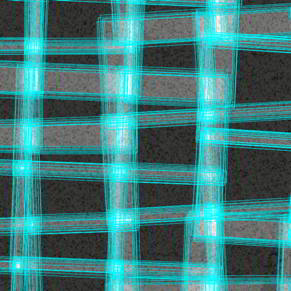}
\end{minipage}
\caption{A detail of the data coverage in the EN1 field at 3.6$\,\mu m$. The image is
  $12''$ across. The \emph{cyan} lines indicate the array edges of the
  individual input images and the \emph{greyscale} underneath shows the mosaic coverage
  depth, ranging from 12 to 35 frames. The non-uniform depth of coverage seen
above is reflected in the cumulative distribution function of Fig~\ref{fig:cdf_cov}.}
\vspace{0.2cm}
\label{fig:epochs}
\end{figure}

Each field was observed in 
two distinct epochs, with the difference in time between the two epochs
ranging from a few days to several months\footnote{When the scheduling
    gap was in months, the AORs were redesigned at the time of the
    observations in order to maintain proper alignment of the tiles.}. This
allows us to reject asteroids, and also  
gives a better photometric accuracy by ensuring that most objects appear
in very different places on the array in the two sets of observations. It arises
from both a half-array offset in array coordinated between the two epochs 
and the fact that the time difference between the execution of each epoch results
in a difference in the field rotation, and hence a different map grid for the 
two different epochs (Figure~\ref{fig:epochs} shows the coverage of the
  two different epochs of observation in the EN1 field).

\begin{table*}
\centering
\caption{Observing log and IRAC Instrument settings for the SERVS fields}
\begin{tabular}{lclcccc}
\hline
\hline
Field and epoch & Dates observed &IRAC&{\em Spitzer}& Array T     & [3.6] bias & [4.5] bias \\
                &                &  Campaign(s)               &ID    & (K) &
(mV)      & (mV)       \\
\hline
EN1 epoch 1& 2009-07-28 to 2009-08-01 & PC1   & 61050        &31              & 450         & 450\\
EN1 epoch 2& 2009-08-02 to 2009-08-05 & PC1   & 61050        &31              & 450 & 450\\
ES1 epoch 1& 2009-08-06 to 2009-08-11 & PC1   & 61051        &31              & 450 & 450\\
ES1 epoch 2& 2009-08-14 to 2009-08-18 & PC2   & 61051        &$\approx 29^*$  & 450 & 450\\
Lockman epoch 1 & 2009-12-10 to 2009-12-21 & PC10,PC11 & 61053      &28.7            & 500 & 500\\
Lockman epoch 2 & 2009-12-25 to 2010-01-04 & PC11,PC12 & 61053      &28.7            & 500 & 500\\
CDFS epoch 1    & 2010-11-01 to 2010-11-13 & PC33,PC34    & 61052        &   28.7          &  500  & 500 \\
CDFS epoch 2$^{\dag}$& 2009-10-13 to 2009-10-28 & PC6,PC7 & 61052 & 28.7 & 500 & 500\\
XMM-LSS epoch 1 & 2010-10-01 to 2010-10-18 &  PC31,PC32   & 60024       & 28.7 & 500 & 500\\
XMM-LSS epoch 2 & 2011-02-19 to 2011-03-06 &   PC42  & 60024       & 28.7 & 500 & 500\\
\hline
\multicolumn{7}{l}{\scriptsize 
$^*$ The array temperatures were allowed to float
  in campaign PC2.}\\
\multicolumn{7}{l}{\scriptsize 
$^{\dag}$ Epoch 2 of CDFS was observed before epoch 1, which proved to be unschedulable 
in its originally planned slot.}
\end{tabular}
\label{table:servsfields}
\vspace{0.5cm}
\end{table*}

Toward the end of the IRAC warm instrument characterization (IWIC), several
tests were performed on variations of the SERVS AORs to establish
which observation strategy was optimal. Three strategies were tested, all
using the small cycling dither pattern, which allows for good coverage while
ensuring that objects are shifted by a minimum of several arcseconds between 
observations.  
The three strategies considered were: (1) two epochs of three dithered 200s
frames, (2) two epochs of three dithered pairs of repeated 100s frames, and
(3) two epochs of six 100s frames. In theory, strategy (1) is the most
efficient and should result in a lower read-noise contribution. However, in
practice, artifacts from bright stars were strong, reducing the effective
area, and the radiation hit (hereafter rad hit) numbers were high (each
array receives approximately 1.5 hits per second, each affecting on average
two pixels, as detailed in the IRAC instrument handbook\footnote{The IRAC instrument handbook
    can be found at \url{http://irsa.ipac.caltech.edu/data/SPITZER/
      docs/irac/iracinstrumenthandbook/}}), resulting in a few 
rad hits leaking through into the final mosaic. There was also no measurable
improvement in noise level compared with the other two options, which used 100s
frames. Option (2) was almost as efficient as option (1), as only a 
fraction of a second was added to the overheads (the 200s frames have a longer
readout time than the 100s frames), but
image persistence effects were significant. 
Option (3) was therefore adopted,
resulting in a very robust survey at the expense of only $\approx 3$\% of
extra observing time. 

The performance of the IRAC camera (both optically, in terms of PSF
and array distortion, and in terms of sensitivity) was
similar to cryogenic performance, hence the survey design was not
modified because of array temperature changes. The sensitivity of the
[3.6] band was affected at the 7\% level by a change in the array bias 
between the taking of the early and later SERVS fields (see details of the
calibration issues in Section~\ref{subsec:calib}), but this variation
was not seen as significant enough to warrant a change in survey strategy. 

The mapping strategy used the small cycling dither pattern, which ensured
full coverage with our map spacing of $280''$. However each epoch of a SERVS
field takes long enough to observe that the field rotation changes
significantly between the start and end of a single epoch of observations. 
Therefore it needed to be robust against the $7^{\circ}$ field rotation between
AORs expected in an $\approx 10$ day window in most SERVS fields. To allow
for this, and to allow for accurate filling out of fixed field geometries,
the SERVS AORs were kept relatively small 
($3\times 3$ maps of $5' \times 5'$ frames) and spaced closely enough to ensure
overlap for the largest expected field rotation. The small AORs also had the
advantage of being easier to schedule, allowing the placement of downlinks
and the insertion of short non-SERVS observations. 

The total observing time for EN1, ES1, Lockman, CDFS and XMM-LSS were 153.4h,
231.8h, 354.5h, 319.1h 323.2h,
respectively. The mean
integration time per pixel of the resulting SERVS mosaics is close to the
design depth at $\approx 1200$s. There are, however, both regions of
significantly deeper data where AORs and map dithers overlap, and shallower
areas, particularly around the edges, or where one epoch is affected by
scattered light from a field star. Uniformity of coverage of the five SERVS
fields is discussed in detail in Section~\ref{subsec:imcov}.

\vspace{0.2cm}
\section{Data processing}
\label{sec:processing}

SERVS data are available from the \emph{Spitzer}
Heritage Archive\footnote{\url{http://sha.ipac.caltech.edu/applications/Spitzer/SHA/}}. 
SERVS mosaics (image, coverage, uncertainty and mask mosaics) and catalogs,
including ancillary data at other wavelengths taken as part of SERVS will be
made available to the community during the summer of 2012, ultimately through
the Infrared Science Archive (IRSA). Catalogs containing the full data set of
ancillary data will be described in detail in Vaccari et al. 2012., in preparation 

\subsection{Image post-processing and mosaics}
\label{subsec:image_post}

Data processing begins with the Basic Calibrated Data (BCD) image, produced by the 
{\em Spitzer} Science Center\footnote{\url{http://ssc.spitzer.caltech.edu/}}
(SSC). These images have been dark-subtracted, flat-fielded, and have had
astrometric and photometric calibration applied. 
A pipeline\footnote{\url{http://irsa.ipac.caltech.edu/data/SPITZER/SWIRE/}} originally used for processing SWIRE data was improved and applied
to the SERVS data to further clean the 
frames of artifacts. Specifically, this pipeline fixed an artifact called
``{\it column pulldown}''\footnote{The \emph{column pulldown} effect, which manifests
  in the slow read direction (columns) of the detectors at 3.6 and $4.5\mu$m, is
  a depression in the zero-level of the column.} found near bright stars, and
also corrected interframe bias offsets  
by setting the background equal to that of a COBE-based model of the zodiacal background
(the dominant background at these wavelengths). Due to the inability to use
the IRAC shutter$^2$, all IRAC data suffer from a variable instrument bias level known as
the ``{\it first-frame effect}'' which varies depending on the recent
detector history. Thus no measurement of the true infrared background, nor of
any spatial structure within the background larger than the array size of {5'}, is possible.  

\begin{figure}[ht]
\vspace{0.2cm}
\begin{minipage}[b]{1\linewidth}
\centering
\begin{tabular}{cc}
\includegraphics[scale=0.7]{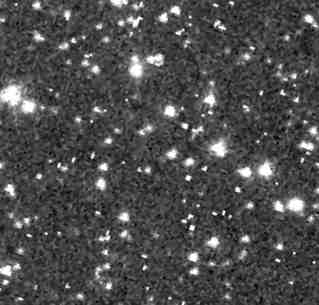} &
\hspace{-0.33cm}
\includegraphics[scale=0.69]{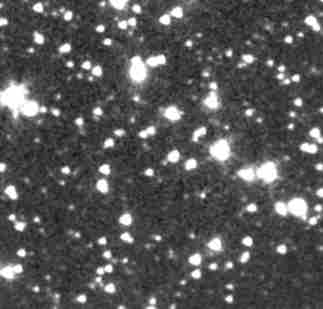}
\end{tabular}
\end{minipage}
\caption{\emph{Left}: An example of a SERVS reduced single 100s
  exposure frame at [3.6]. The frame size is $\sim 5'\times
  5'$. \emph{Right}: the final co-added [3.6] image.\vspace{0.1cm}}
\label{fig:singlecoadd}
\end{figure}

The data were co-added (see Figure~\ref{fig:singlecoadd}) using the {\sc
  Mopex}\footnote{{\sc Mopex} and its associated documentation can be
  obtained at
  \url{http://irsa.ipac.caltech.edu/data/SPITZER/
    docs/dataanalysistools/tools/mopex/}} package available from the SSC
(parameters used are listed in Appendix~\ref{app:sextractor}). All
the data from a 
single field were co-added onto a single frame; the two different wavelengths 
are reprojected to the same astrometric projection so that their pixels align one-to-one.
The data are reprojected with a linear interpolation onto a pixel scale of $0.6''$, 
providing marginal sampling at $3.6\, \mu$m. The multiple dithers allow at least some 
recovery from the severe undersampling of the IRAC camera at these
wavelengths. 

\begin{figure}[ht]
\vspace{0.2cm}
\begin{minipage}[b]{1\linewidth}
\centering
\begin{tabular}{cc}
\hspace{-0.2cm}
\includegraphics[scale=0.86]{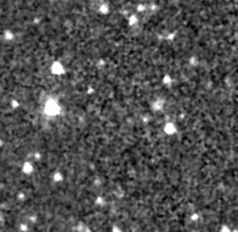} &
\hspace{-0.4cm}
\includegraphics[trim=0mm 0mm 0mm 0mm, clip=true, scale=0.83]{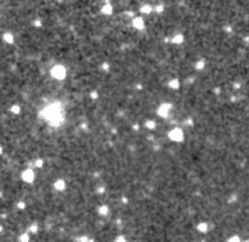}\\
\hspace{-0.17cm}
\includegraphics[scale=0.795]{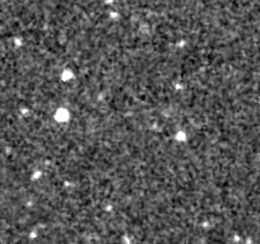} &
\hspace{-0.35cm}
\includegraphics[trim=0mm 0mm 0mm 0mm, clip=true, scale=0.8]{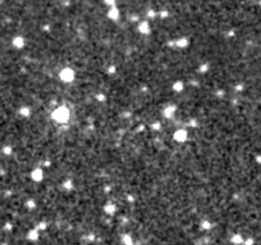}\\
\end{tabular}
\end{minipage}
\vspace{0.3cm}
\caption{A cutout of the EN1 field at [3.6] (\emph{top panel}) \& [4.5]
  (\emph{bottom panel}), as imaged by SWIRE (\emph{left}) \& SERVS
  (\emph{right}). The difference in depth between the two surveys can clearly
  be seen here. 
}
\label{fig:servs_swire_comp}
\end{figure}

The depth reached by the SERVS observations can easily be put in perspective
when comparing SWIRE and SERVS cutouts of a similar region of sky
(\emph{left} and \emph{right} respectively in Figure~\ref{fig:servs_swire_comp}). With a depth
of $3.7\, \mu$Jy at [3.6] and $7.4\, \mu$Jy at [4.5], SWIRE is limited to $z
\approx 1.5$ for $L^{*}$ galaxies, whereas SERVS can detect these galaxies up
to $z \sim 4$ (see also Figure~\ref{fig:detlum}). 

\begin{figure*}
\centering
\includegraphics[scale=0.65]{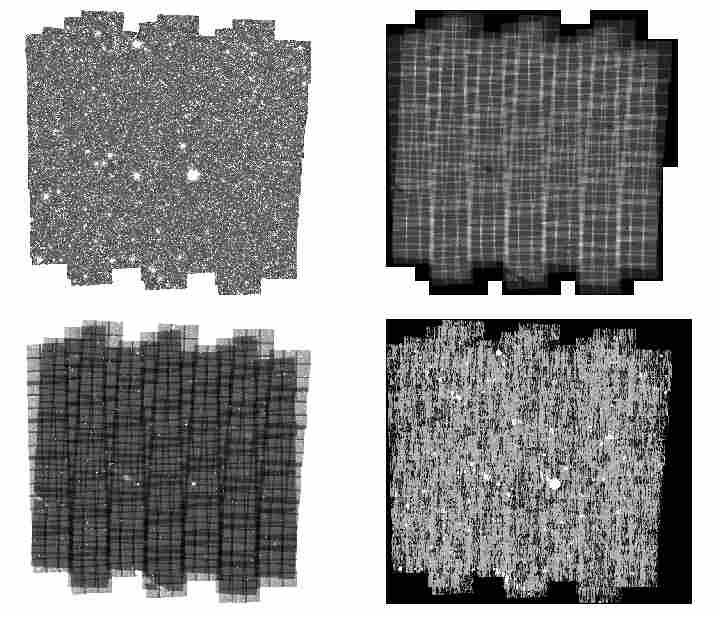}
\caption{SERVS final set of mosaics for the EN1 field at $3.6\mu m$ (\emph{top left}: image
  mosaic, \emph{top right}: coverage mosaic, \emph{bottom left}: uncertainty
  mosaic, \emph{bottom right}: mask mosaic).} 
\vspace{0.3cm}
\label{fig:all4mosaics}
\end{figure*}

Large mosaics containing all the SERVS epochs of observation in each field
were created from the co-added images, along with the uncertainty, coverage
and mask images (see Figure~\ref{fig:all4mosaics}). The resulting mosaics
cover areas of $\sim \,$2, 3, 4, 4.5 and 4.5 deg$^2$ for EN1, ES1, Lockman, CDFS
and XMM-LSS, respectively (see Table~\ref{table:servsgeometry} for details about
the geometry of the five fields). 

\subsection{Calibration issues}
\label{subsec:calib}

The initial observations (all of EN1 and the first half of
ES1) were made before the IRAC detectors were stabilized at their
current operating temperature of 28.7K, which is now the norm for the 
{\em Spitzer} post-cryogenic (i.e. ``warm'') mission (see
Table~\ref{table:servsfields}). Instead, these data were taken at a
controlled temperature of 31K. The second half of 
ES1 was taken during a time when the
detector was cooling from 31K to the final temperature of 28.7K, and
was not under active temperature control. In addition, during this
time period the detector biases for both arrays were
adjusted. These changes in temperature and detector operating
parameters resulted in changes to both the instrument calibration and
its noise properties. These were only measurable in the [3.6] band, 
resulting in an $\approx 7$\% increase in the noise in EN1 and ES1 
at [3.6] compared with the remainder of the survey. 

IRAC is calibrated based on dedicated calibration observations
collected during science operations. These data are typically collected
on timescales of years. The \emph{Spitzer} cryogenic mission, as well as the
nominal warm mission, are extremely well-calibrated. However, during
this transition period only a small amount of calibration data could
be taken, as conditions were constantly changing. The \emph{Spitzer} Science
Center provided an initial calibration, which was reliable to a few
percent. Photometry from SERVS was compared with that of SWIRE on an
object-by-object basis, and small multiplicative offsets at the few
percent level were found. These calibration errors were fixed at the
catalog level  (see Section~\ref{sec:catalogs}) by applying
multiplicative factors derived from comparison between 
sources detected both in SWIRE and SERVS. Calibration is therefore the
same as that of the SWIRE data.

\subsection{Uniformity of coverage}
\label{subsec:imcov}

\begin{figure*}
\begin{tabular}{cc}
\hspace{0.3cm}
\includegraphics[trim=0cm 0cm 0cm 0cm,clip=true,scale=0.45]{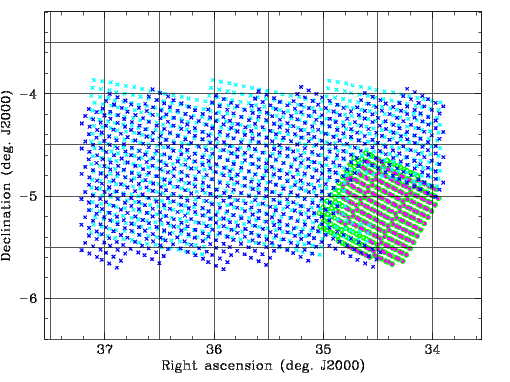}&
\includegraphics[trim=0cm 0.2cm 0cm 0cm,clip=true,width=9cm,height=6.8cm]{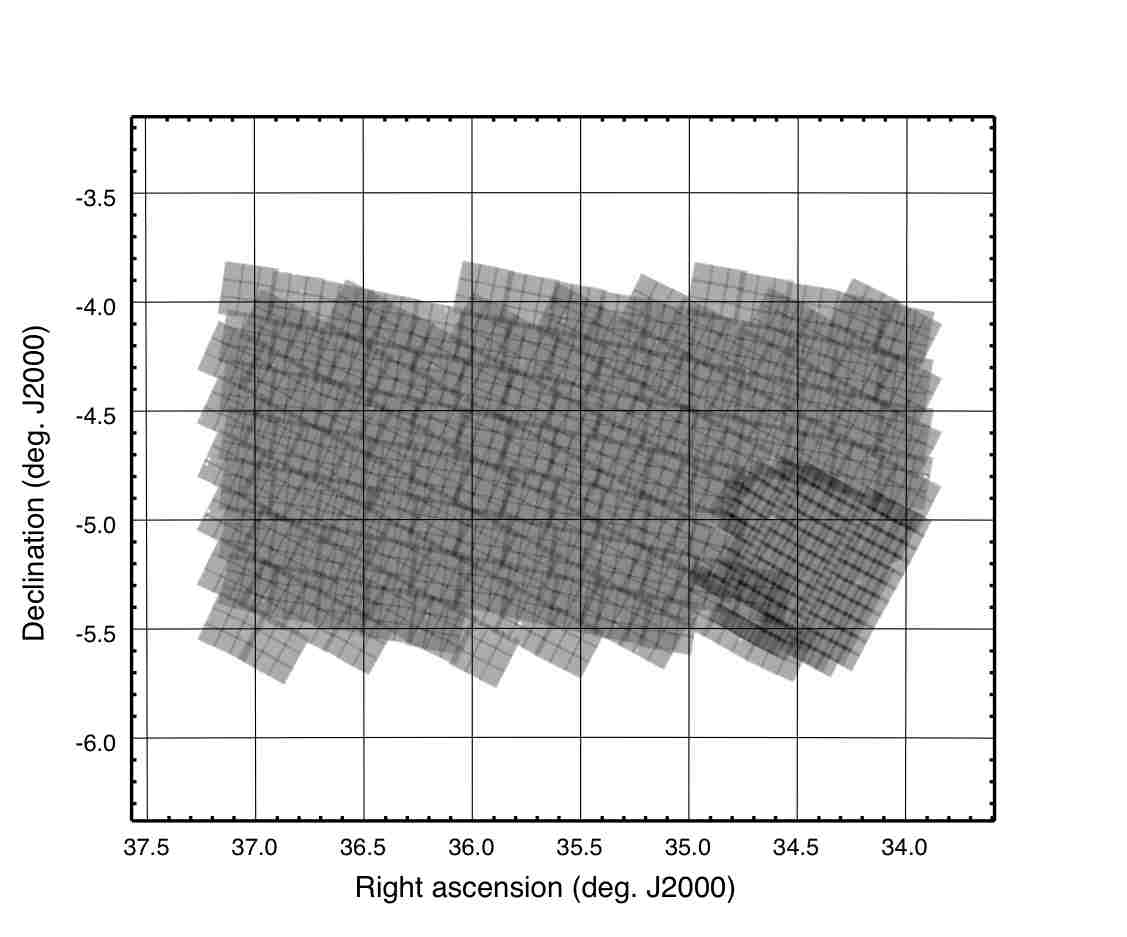}\\
\end{tabular}
\caption{
\emph{Left}: XMM-LSS single frame centers, showing epoch 1 in \emph{cyan} and
  epoch 2 in \emph{blue} crosses (see Table~\ref{table:servsfields}). A
  grid extrapolating the XMM-LSS epoch2 geometry was overlaid onto the SpUDS
  field frames (\emph{in green open circles}) and used to select the relevant SpUDS
  frames to co-add (\emph{magenta solid points}) to complete the XMM-LSS field. 
\emph{Right}: Coverage map in equatorial coordinates of the XMM-LSS field at [3.6]. Note the
addition of the SpUDS data between $ 35 \leq \alpha \leq 34$ \& $-5.7 \leq
\delta \leq -4.7$ to patch the XMM-LSS field. The rotation in tiling induced by
the choice of two different epochs of observation (of six $100\,$s frames each,
as discussed in Section~\ref{subsec:AOR_design}) is also clearly visible
here. Coverage is not completely uniform throughout the field and data can be deeper, especially
at the overlap of the SERVS (warm mission) \& SpUDS (cryogenic) surveys (see
Section~\ref{subsec:imcov} for details).}  
\vspace{0.5cm}
\label{fig:covmaps}
\end{figure*}

Certain areas of the original SERVS footprint were already covered by several
previous {\em Spitzer} surveys, such as SpUDS for XMM-LSS and SIMPLE/GOODS for CDFS (see
Section~\ref{subsec:fieldselection} for details). These regions were
deliberately avoided during the SERVS observation campaign to save
observation time. Since the optical properties of the camera did not 
change between the cryogenic and warm missions and since all the
\emph{Spitzer} data was combined using the same
{\sc Mopex} pipeline, previous survey  
frames were subsequently merged with the SERVS data at co-addition and 
patched onto the final mosaics. Given the higher depth
of the previously existing smaller \emph{Spitzer} surveys, it was always
possible to reasonably match the depths of the SERVS observations in these
areas by selecting the right number of single frames to co-add. In addition,
since the archival data are only a minimal fraction of the survey ($\sim 12\,\%$
of XMM-LSS and $\sim 3\,\%$ of CDFS, none in the other fields), these minor differences in coverage
between archival and SERVS data are not an issue.  

The selection of archival data was focused on filling the missing areas as best
  as possible and thus not refined to be necessarily  chosen from
different dates of observations, hence some contamination by transient sources
on timescales of hours is possible. Some artifacts such as
\emph{muxbleed}\footnote{The \emph{muxbleed} effect appears as a series of bright
  pixels along the fast read direction (horizontal in  array coordinates),
  and may extend the entire width of the array ($5'$).} were
present in the cryogenic data and not in SERVS, however after processing, all
artifacts were removed and did not impact the mosaicking.  
A grid extending over the missing data was set to match the SERVS frame
centers (spacing and orientation) as closely as possible. 
The closest SpUDS (or SIMPLE/GOODS) frames to these grid centers were then
automatically selected and a subsample of those was chosen to best cover the
area and match the SERVS depth (the left panel of Figure~\ref{fig:covmaps}
shows the XMM-LSS field frame selection process as an example). The resulting
coverage map for the XMM-LSS field is shown in the right panel of
Figure~\ref{fig:covmaps} as an example. 

\begin{figure}
\resizebox{\hsize}{!}{\includegraphics[angle=-90,scale=0.6]{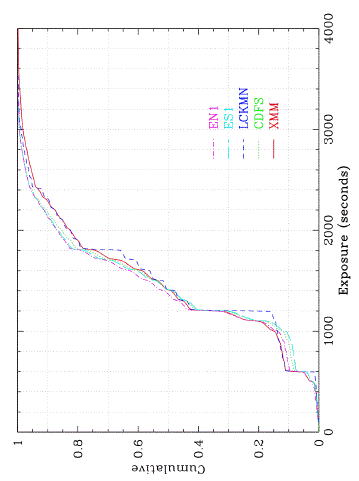}}
\vspace{0.2cm}
\caption{Cumulative distribution function of the coverage maps for all five
  fields, in terms of exposure time (in seconds). The curve represents the 
  fraction of pixels with that integration time or lesser.} 
\label{fig:cdf_cov}
\end{figure}

Coverage is thus not completely uniform throughout the fields, though averaging
$\sim 1400\,$s of exposure time over all five fields (see 
Figure~\ref{fig:cdf_cov}). By design some overlap
between AORs was allowed and some areas in a field may have a higher coverage
(e.g. at the AOR intersections), for example totalling up to 
$\sim 4700\,$s in
Lockman. Although great care was taken to minimize variations in the coverage
throughout a field, the exposure time can rise up to $\sim 5000\,$s (e.g. in
XMM-LSS) when the added ancillary data intersects both SERVS epochs (mostly
around the edges of the SERVS/ancillary data, as is obvious in the right panel of
Figure~\ref{fig:covmaps}). However these areas of higher coverage
  represent a very small portion of the fields and do not have an impact on
  the overall uniformity, given the already intrinsic nonuniform nature of
  the SERVS coverage due to the original tiling design. The higher depths of 
  the observations at the frames overlaps or due to archival data
  overlapping the SERVS frames do not result in any impact to the survey
  reliability or completeness when a minimum flux
  selection based on the SERVS data is used.

\subsection{Image masks}
\label{subsec:immasks}

Bright source artifacts (straylight, column pulldown, latents) are a
significant problem with {\em Spitzer} data. Most of these artifacts are discussed
in more detail in the IRAC instrument handbook$^4$, available from the SSC, as well as
in \cite{Surace+04} and \cite{Surace+05}$^7$. In order to flag for these artifacts, mask images
are created for each of the five fields. Flags bit values attributed to each
significant artifacts are listed in Table~\ref{table:maskbits}, as well as in
the headers of the mask FITS files.

\begin{table}
\begin{center}
\begin{minipage}[b]{1\linewidth}
\caption{Lookup table for the mask image bit values}
\label{table:maskbits}
\begin{tabular}{lcl}
\hline
\hline
Bit number &  Bit value    & \hspace{2cm} Flag  \\
\hline
BIT00   = &  1 & Overall data quality              \\
BIT01   = &  2 & Set if pixel contains rad hit      \\     
BIT02   = &  4 & Set if optical ghost present      \\
BIT03   = &  8 & Set if stray light present        \\
BIT04   = &  16 & Set if saturation donut           \\
BIT05   = &  32 & Set if pixel contains latent image\\
BIT06   = &  64 & Set if pixel is saturated         \\        
BIT07   = &  128 & Set if column pulldown present    \\
BIT08   = &  512 & Set if bright star is present \\ 
\tableline
\multicolumn{3}{l}{\scriptsize {\it Notes}: Flags are described in more detail in the text in
  Section~\ref{subsec:immasks}}
\end{tabular}
\end{minipage}
\end{center}
\end{table}

\begin{figure}
\begin{center}
\includegraphics[scale=0.8]{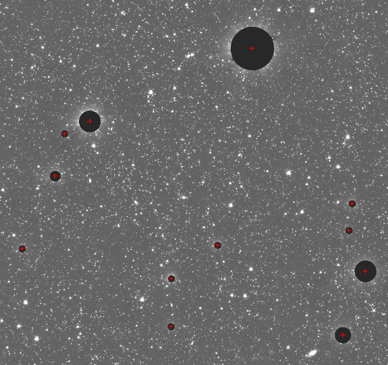}
\vspace{0.4cm} 
\caption{Bright star mask cutout of a portion of the Lockman field. \emph{Black
  circles} show the masked regions around the cross-identified 2MASS bright stars centers
  (\emph{red crosses}) in the field. Radii are proportional to 2MASS K-band
  brightness and shown in Table~\ref{tab:bs_flag}.}
\label{fig:lckmn_bs_cutout}
\end{center}
\vspace{0.3cm}
\end{figure}

In addition to these known {\em Spitzer} image artifacts, saturated stars
are common across the deep and wide SERVS fields and thus need to be flagged
appropriately. Indeed very bright stars do not have reliably extracted fluxes
in SERVS, and may occasionally be split into multiple fainter sources, or are
saturated, so no flux can be measured accurately; erroneous detections in
the wings of the PSF can also cause artifacts. As a result, a safe radius has to be
set and flagged around those bright stars. 
Luckily, any object
triggering bright star artifacts in the SERVS data is easily detected by the
Two Micron All Sky Survey (2MASS,\citealt{Skrutskie2006}), which has 
reliable fluxes even for very bright objects. The positions and K-band
magnitudes of bright stars are downloaded from the 2MASS Point Source Catalog
catalog in the Vizier database\footnote{The Vizier database is available at
  http://vizier.u-strasbg.fr/}. Magnitudes are thus converted 
into radii according to Table~\ref{tab:bs_flag}. The radii are taken from
SWIRE \citep{Surace+04} and increased to take into account the deeper
exposures of SERVS.
To reduce memory requirements on such large images, subsets of the mosaics
were cut around each bright star, flagged and then 
reembedded onto the final mask mosaic. The resulting bright star flag
masks a circular region around each bright star in the fields. A part of the
Lockman field bright star mask is shown in Figure~\ref{fig:lckmn_bs_cutout}.  

\begin{table}
\vspace{0.2cm}
\begin{center}
\caption{Bright Object flagging for the mask images}
\label{tab:bs_flag}
\begin{tabular}{lc}
\hline
\hline
2MASS--K mag. range       & radius ($''$) \\
\hline
$> 12.0$ &   0    \\
$10.0 - 12.0$ &  15   \\
$8.0 - 10.0$ &   20 \\
$7.5 - 8.0$ &   30 \\
$6.5 - 7.5$ &   45  \\
$5.0 - 6.5$ &   60 \\
$< 5.0$ &  120  \\
\hline
\end{tabular}
\end{center}
\scriptsize 
{\it Notes}: The radii are taken from SWIRE
\citep{Surace+04} and increased to take into account the deeper exposures
of SERVS.
\vspace{0.3cm}
\end{table}

\vspace{0.5cm}
\section{Catalogs}
\label{sec:catalogs}

\subsection{Catalogs extraction and calibration}
\label{subsec:catextrac}

Catalogs were made using {\sc SExtractor} \citep{BA1996}. Two sets of
catalogs were produced for each field (the parameters used in the source
extraction are given in Appendix~\ref{app:sextractor}). The first set is based on extractions
from the [3.6] and [4.5] bands, separately. A second set is then created when
the two catalogs are merged, and only detections common to both are
retained. This results in a high-reliability catalog that is used to combine
with other data sets. For rare objects searches, we recommend using the
high-reliability catalogs. The single-band catalog may be used for
bandmerging with other data (e.g. UKIDSS/DXS data), or for statistical studies. 

SExtractor aperture fluxes are computed within radii of $1.4'', 1.9'', 2.9'', 4.1''$ and
$5.8''$ and corrected using the aperture correction factors derived for SWIRE
DR2/3\citep{Surace+04} and reported in Table~\ref{tab:apsizecorr}. The
IRAC instrument has a calibration tied to standard 
stars as measured in a fiducial $12"$ radius aperture. This aperture is
non-ideal for faint-source extraction, so the fluxes are measured in smaller
apertures and suitable aperture corrections are applied.

A fraction of the SERVS data was taken prior to the final temperature
stabilization of IRAC, and prior to the selection of the final array biases 
(affecting the EN1 and ES1 fields, see
Table~\ref{table:servsfields}). During this 
``floating temperature'' period, the IRAC 
calibration drifted, with changes to both the overall gain and the detector
linearization, as discussed Section~\ref{subsec:calib}. As a result, the
images have overall calibration errors at the 
level of a few percent. In addition, the other fields show small, but
noticeable calibration 
differences compared to the cryogenic [3.6] SWIRE data. 
These calibration errors were fixed at the catalog
level by applying multiplicative factors
derived from comparison between 
sources detected both in SWIRE and SERVS
($f_{\rm{[3.6]SWIRE}}/f_{\rm{[3.6]SERVS}}=1.07$ for EN1 \& ES1 and 1.02
for Lockman, CDFS and XMM-LSS; at 
[4.5] the correction factors are very
close to unity, and no corrections were applied). 
SERVS calibration is thus tied to that of the SWIRE data.

\begin{table}
\begin{center}
\caption{Aperture sizes and corrections}
\label{tab:apsizecorr}
\begin{tabular}{cccc}
\hline
\hline
Aperture & Aperture radii  & [3.6] correction & [4.5] correction\\
  number              &  (arcsec)          &    (arcsec)     &  (arcsec)  \\
\hline
ap1 & 1.4    &  0.585  &  0.569   \\
ap2 & 1.9    &  0.736  &  0.716   \\
ap3 & 2.9    &  0.87   &  0.87    \\
ap4 & 4.1    &  0.92   &  0.905   \\
ap5 & 5.8    &  0.96   &  0.95    \\\hline
\end{tabular}
\end{center}
\noindent 
\scriptsize 
{\it Notes}: Aperture sizes and corrections were derived for SWIRE
\cite{Surace+04}. More details can be found therein. 
\vspace{0.5cm}
\end{table}

\subsection{Number counts and completeness}
\label{subsec:data_analysis}

\begin{figure}
\centering
\begin{tabular}{c}
\includegraphics[scale=0.33]{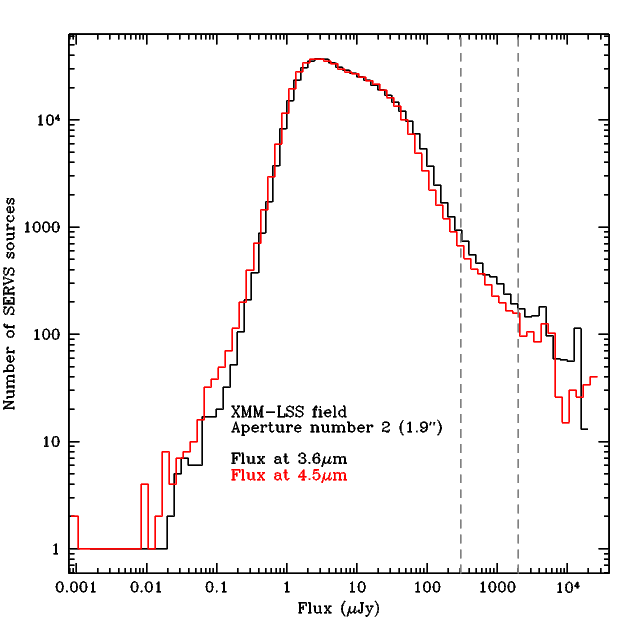}\\
\includegraphics[scale=0.33]{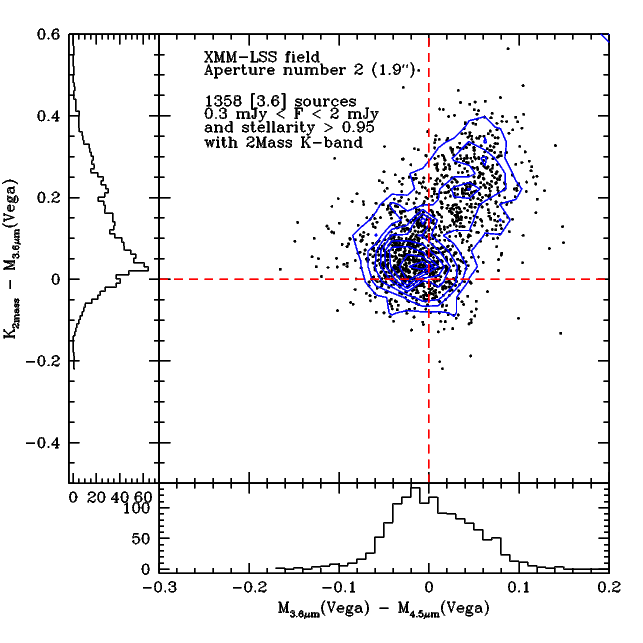}
\end{tabular}
\caption{\emph{Top}: SERVS number counts versus flux at [3.6] (\emph{black}
  histogram) and
  [4.5] (\emph{red} histogram) for the XMM-LSS field. Grey dashed lines show the selection of
  sources used in the right plot (with fluxes as $0.3\, < f < 2\,$mJy). 
  \emph{Bottom}: Color-color plot showing $K_{\rm{2MASS}} - M_{3.6\mu
    m}$(Vega) versus $M_{3.6\mu m}{\rm (Vega)}-M_{4.5\mu m}$(Vega) for sources
  within the flux range defined by the grey dashed lines above, plus a cut
  in stellarity index $>0.95$ and the existence of a 2MASS K-band
  measurement as an additional constraint. \emph{Red dashed} lines help
  pinpoint the location of the (0,0) point in this diagram. Similar plots for
  the EN1, ES1, Lockman \& CFDS fields are shown in
  Appendix~\ref{app:fourfieldsnumcounts}.}  
\vspace{0.2cm}
\label{fig:xmmnumcounts}
\end{figure}

Simple number counts were derived from the extracted catalogs. The SERVS
source counts for the XMM-LSS field is provided as an example in the left part of
Figure~\ref{fig:xmmnumcounts} (the remaining four fields are shown in
Appendix~\ref{app:fourfieldsnumcounts}). Several features are visible in this
plot. Fundamentally, the observed counts present as a power-law. This is
shallow at the bright end, primarily due to the presence of bright
stars. There is a known break in the power-law index near 100\,$\mu$Jy
\citep{Glazebrook+94}. At the faintest levels, the turnover is a result
of the increasing incompleteness of the survey. It is clear that the SERVS
completeness level catastrophically drops near 2--3\,$\mu$Jy. 
As a consistency check, a sample of bright stars (SERVS fluxes within $0.3< f <
2\,$mJy and stellarity index $> 0.95$) was selected. Cross-identifying with
the 2MASS catalog, K-band measurements were used to compare the deviation of
SERVS colors with respect to the 2MASS ones. A color-color plot of
$K_{\rm{2MASS}} - M_{3.6\mu m}$(Vega) versus $M_{3.6\mu m}{\rm (Vega)} -
M_{4.5\mu m}$(Vega) for the XMM-LSS field, shown in the right part of
Figure~\ref{fig:xmmnumcounts}, confirms that SERVS is consistent with 2MASS.

SERVS detects $\sim 100,000$ sources per square degree and, with $\sim 40$ beams
per source, approaches the classical definition of where source confusion
becomes important. The effective depth of SERVS is thus affected by confusion
noise, which makes the definition of survey depth dependent on the experiment
one wishes to carry out. For a detection experiment on point-source objects
detected in another waveband with a well-determined position (positional
uncertainty much lower than the SERVS beam of $\approx\,$2$''$), the
appropriate number, denoted $\sigma_{\rm pp}$, is determined from the
pixel-to-pixel variance in a source-free region of a single
basic calibrated data (assuming it scales as the square root of the
coverage and that extraction is carried out by source fitting; i.e.\
assuming 7.0 and 7.2 noise pixels in [3.6] and [4.5], respectively).
Another measure of the noise, which includes some
contribution from confusion noise, is obtained from ``empty-aperture''
measurements, where the standard deviation of fluxes in object-free
apertures in the final mosaic is measured directly.
For our fields, this
measurement was made in $3.8''$ diameter apertures (SWIRE aperture 2), and is
denoted $\sigma_{\rm ap}$. SWIRE aperture 2 is recommended by the SWIRE team
as the most stable aperture for photometry since most faint IRAC sources are
slightly resolved at the $\approx 1-2''$ level. Survey depths for the various
measurements are summarized in Table~\ref{tab:surveydepths}.  

\begin{table}
\vspace{0.2cm}
\caption{Approximate survey depth and completeness in the SERVS fields}
\vspace{-0.2cm}
\begin{center}
\begin{tabular}{lcc}
\hline
\hline
Measurement       & [3.6] value & [4.5] value\\
                  &($\mu$Jy)    &  ($\mu$Jy)  \\
\hline
5\,$\sigma_{\rm pp}$ &       1.3  &     1.5   \\
5\,$\sigma_{\rm ap}$ &       1.9  & 2.2       \\
$S_{\rm 50c}^{\dag}$ &      $4.0,\;3.0$      &  $3.5,\;3.5$ \\
$S_{\rm 80c}^{\dag}$ &      $7,\;5$          &  $5,\;5$      \\
 \hline
\end{tabular}
\end{center}
\scriptsize 
$^{\dag}$ The two completeness
levels shown correspond to the the two-band detection catalog and the
single-band catalog.
\newline {\it Notes}: EN1 and ES1 are about 7\% noisier than the other fields in [3.6]
due to different detector settings used early in the warm mission
(Table~\ref{table:servsfields}).
\vspace{0.2cm}
\label{tab:surveydepths}
\end{table}

\begin{figure}
\vspace{0.1cm}
\resizebox{\hsize}{!}{\includegraphics[scale=0.65]{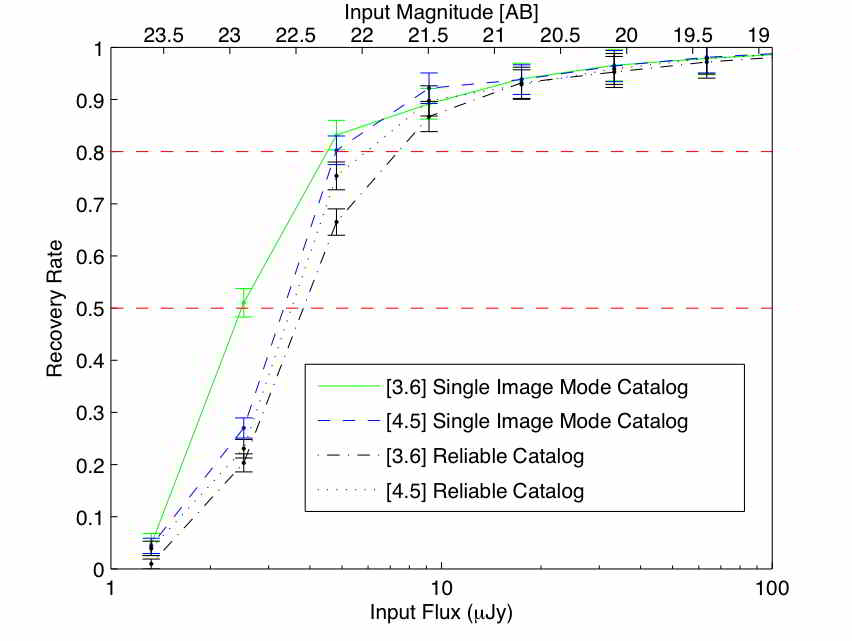}}
\caption{Completeness plots for Lockman Hole at [3.6] shown in green, [4.5]
  shown in blue, and the reliable dual-band detection catalogs shown in
  black. Using the techniques of \cite{Chary+04} and \cite{Lacy+05},
  $\sim$10,000 model galaxies were simulated and placed in the mosaics. The
  sources were extracted using the same pipeline as the catalogs and the
  completeness was estimated as the recovery rate of the simulated model
  galaxies.} 
\label{fig:en1comp}
\vspace{0.3cm}
\end{figure}

Finally, for survey work, the completeness limits at 50 and 80\% ($S_{\rm
  c50}$ and $S_{\rm c80}$), give a good
indication of the depths that are usable for global survey properties such as
source counts. Following the techniques of \cite{Chary+04} and
\cite{Lacy+05}, we simulated $\sim$10,000 model galaxies, distributed them in
the reduced mosaics, and extracted them using the same pipeline that created
the catalogs. The recovery rate of the model galaxies was used as our
completeness indicator and suggests that $\approx$\,50\% completeness is
reached at a flux density of $\approx$\,$2-3$\,$\mu$Jy in the single band catalogs
at both [3.6] and [4.5], due to a combination of signal-to-noise ratio and source
confusion (see Figure~\ref{fig:en1comp}).  More details of the completeness
and reliability of the SERVS catalogs will be presented by Vaccari et
al.\ (2012, in preparation).

\subsection{Expected detection limits \& redshift distribution}
\label{subsec:detlim}

The SERVS project uses semi-analytic models extensively, both to make
testable predictions of the properties of SERVS galaxies, and to inform our
follow-up strategies in wavebands other than the near-infrared
(e.g. Figure~\ref{fig:depths}). Two different semi-analytic models are
currently being used. 

The first set is based on the \cite{Guo+11} version of
the Munich semi-analytic model for which lightcones\footnote{These lightcones
  are publicly available at http://www.mpa-garching.mpg.de/millennium} were
created and fully described in Henriques et al. (2012). The lightcones contain a wide
range of photometric bands that cover the UV to near-infrared region of the
spectra and allow an exact match to the observed selection criteria. They
also include a choice between fluxes computed using the \cite{BC2003} or the
\cite{Maraston05} stellar populations. The latest have been shown to
reconcile the predicted $K$-band rest-frame luminosity function with
observations at high redshift (\citealt{Henriques+11}, Henriques et al. 2012), for
which IRAC data - such as those obtained with SERVS - have been essential.   
The second set, from van Kampen
et al.\ (2012a, in preparation), includes both the effects of halo-halo and galaxy-galaxy
mergers, and uses GRASIL \citep{Silva+98} to predict spectral energy
distributions (SEDs) from the optical to submillimeter. 

\begin{figure}
\resizebox{\hsize}{!}{\includegraphics{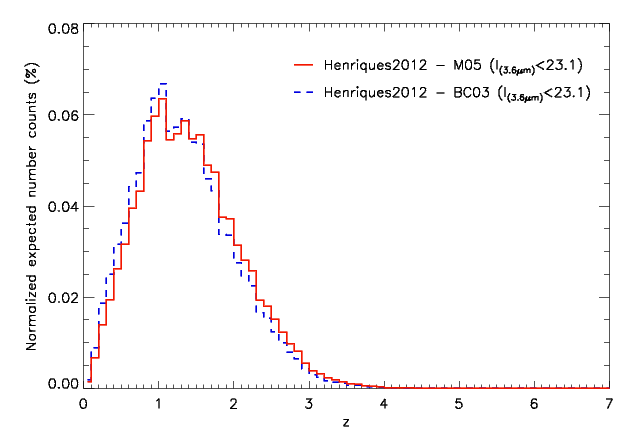}}
\caption{SERVS expected normalized galaxy counts at the detection limit of
  $2\, \mu$Jy (or $m_{[3.6]} < 23.1$). The two histograms 
  correspond to flux limits applied to the [3.6] band either using the
  \cite{Maraston05} or \cite{BC2003} stellar population models (see
  Section~\ref{subsec:detlim}).}
\label{fig:zdist}
\vspace{0.3cm}
\end{figure}

The expected redshift distribution is derived from the simulations and shown in
Figure~\ref{fig:zdist}. A lower flux limit of $2\, \mu$Jy (AB
magnitude of $m_{[3.6]} < 23.1$) for the SERVS survey was used here. The
expected redshift distribution of the SERVS galaxies peaks around $z \sim 1$
and extends to redshifts of $z \sim 3$, with a small fraction of objects
detected up to $z \sim 4$ (for basic comparison purposes, the SWIRE
  photometric redshift distribution can be found in
  \citealt{RR+08}; at the current time, the SERVS photometric redshift 
    distribution has not been derived but will be presented by Pforr et
    al. 2012b, in preparation, as discussed in 
  Section~\ref{subsec:photoz}). Galaxies detected in SERVS will 
  therefore span the epochs 
  where galaxies gain the vast majority of their stellar 
mass. Indeed, \cite{Brown+07} and \cite{Cool+08} estimate that
  $L^*$ galaxies roughly double in mass between $z=0$ and $z\approx\,$1. In
addition, \cite{vanDokkum+10} recently showed that about half the mass of any
given large galaxy is added between $z=0$ and $z=2$, by comparing galaxy
samples at constant number densities. SERVS will be able to extend such
studies out to higher redshifts with good statistics. 

\begin{figure}
\resizebox{\hsize}{!}{\includegraphics{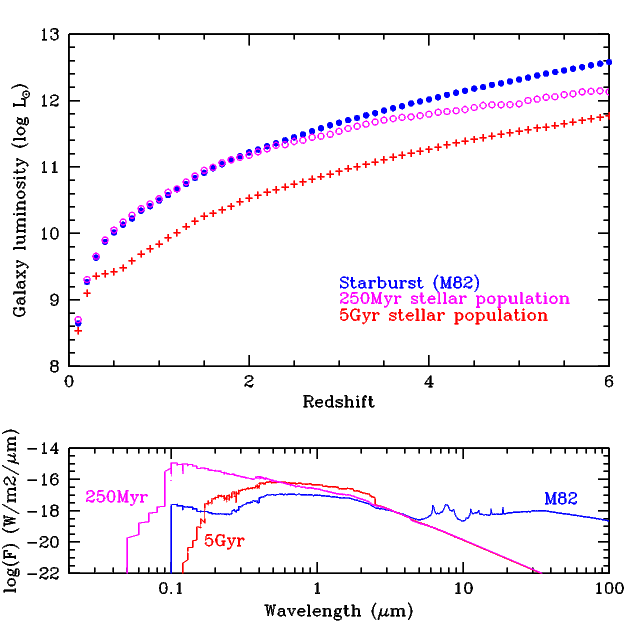}}
\caption{{\it Top panel}: SERVS faintest galaxy to be detected, in units of
  solar luminosities (integrated luminosity from the optical to the far-IR: $0.1\,\mu$m--$1\,$mm)
  at a given redshift. Three Spectral Energy Distribution (SED) 
  models are considered here and shown in the {\it bottom panel}. The 
  three SEDs consist of an M82-like starburst galaxy (in \emph{blue}), a
  250 Myr stellar population (\emph{magenta}) and a 5 Gyr stellar population
  (\emph{red}), scaled to the same $3.6 \mu$m restframe. SERVS can detect $L^{*}$ 
  galaxies out to $z \sim 4$, and 0.1$L^{*}$ out to $z \sim 1$. At redshift
  5, a starburst galaxy of luminosity $\sim 10^{12}\Lsun$ could
  potentially be detected.}
\vspace{0.5cm}
\label{fig:detlum}
\end{figure}

In order to roughly estimate the luminosity of the faintest galaxy SERVS will be able to detect
in the IRAC [3.6] band as a function of redshift, three model SEDs were used: a
starburst similar to M82, a 250 Myr stellar population and a 5 Gyr stellar
population from \cite{Maraston05}. Figure~\ref{fig:detlum} shows the
luminosity of the faintest detectable source from $0.1\,\mu$m to 1\,mm as a
function of redshift, if this source was represented by one of the three SED
models we used. SERVS might be able to detect $L^{*}$ galaxies up to $z \sim
4$ and 0.1$L^{*}$ out to $z \sim 1$. The luminosities of the Maraston SEDs
can easily be converted to stellar masses by multiplying the luminosity of
the 5 Gyr and 250 Myr stellar population models by 1.15865 and 0.00133
respectively. SERVS will ensure the derivation of robust stellar masses 
because it includes the rest-frame near-infrared out to high redshifts. This
coverage is essential to break the degeneracy between star formation history
and dust reddening \citep{Maraston+06}. Photometric redshifts for SERVS
galaxies are also being derived (see section~\ref{subsec:photoz}). 

\vspace{0.5cm}
\section{Ancillary data}
\label{sec:ancillary}

Overlaps with existing and surveys in progress are listed in
Table~\ref{table:ancillarysurveys}. Here we provide a more detailed
description of some of the most significant (in terms of their overlaps with
SERVS) of these surveys.

\begin{table*}
\begin{center}
\caption{Surveys at other wavelengths covering $\stackrel{>}{_{\sim}}$10\% of a SERVS field (all-sky surveys
excepted).}
\label{table:ancillarysurveys}
\begin{tabular}{lcccccc}
\hline
\hline
Field    & X-ray      & Optical       & Near-IR    & Mid-IR   & Far-IR/     &Radio\\
Name     & data       & Data          & Data       & Data     & submm Data  &Data \\
         &            &               &            &          &             &    \\\hline\hline
EN1 & N04$^{1}$  & SWIRE/INT$^{2}$ & DXS$^{5}$ &SWIRE$^{6}$ & HerMES$^{7}$ L5& G08a$^{9}$\\  
         &            & SDSS$^{3}$      &            &          & S2CLS$^{8}$ & Gr10$^{10}$        \\
         &            &  SpARCS$^{4}$   &            &          &              &    \\\hline
ES1 &F08$^{11}$   & ESIS$^{12}$         & VIDEO$^{14}$  & SWIRE$^{6}$&HerMES$^{7}$ L6 &ATLAS$^{15}$\\
         &            & VOICE$^{13}$, SpARCS$^{4}$   &            &          &              &    \\\hline
Lockman  &W09$^{16}$   & SWIRE/KPNO$+$INT$^{2}$ & DXS$^{5}$ &SWIRE$^{6}$& HerMES$^{7}$ L3,L5& OM08$^{17}$\\
         &            & SDSS$^{3}$          &            &          & S2CLS$^{8}$ &G08b$^{18}$,G10$^{19}$\\
         &            & SpARCS$^{4}$        &            &          &              & I09$^{20}$\\\hline
CDFS     &CDFS$^{21}$  & SWIRE/CTIO$^{22}$&VIDEO$^{14}$& SWIRE$^{6}$  & HerMES$^{7}$ L2,L5& ATLAS$^{15}$ \\
         &            &  GaBoDS$^{23}$  &           & SIMPLE$^{25}$&LABOCA/LESS$^{26}$   &M08$^{27}$\\
         &            & VVDS$^{24}$   &             &              &                &         \\
         &            & VOICE$^{13}$, SpARCS$^{4}$  &            &          &              &    \\\hline
XMM-LSS  &XMM-LSS$^{28}$ & CFHTLS$^{30}$&VVDS$^{31}$ &SWIRE$^{6}$ & HerMES$^{7}$ L5 & VVDS-VLA$^{32}$ \\
         & SXDS$^{29}$   & VVDS$^{24}$ &UDS$^{5}$    &SpUDS$^{25}$& S2CLS$^{8}$ &S06$^{33}$        \\
         &            &   SXDS$^{34}$            & VIDEO$^{14}$&            &              &         \\
         &            &               &DXS$^{5}$    &            & &         \\
\hline\\
\end{tabular}
\end{center}
\scriptsize

{\it Notes}: 
[1] Chandra proposal 6900602 (P.I.\ Nandra); 
[2] \cite{Gonzalez+11};
[3] \cite{Abazajian+09};
[4] DeGroot et al. 2012, in preparation
[5] \cite{Lawrence+07}; 
[6] \cite{Lonsdale+03}, \cite{Surace+05}$^7$; 
[7] \cite{Oliver+12}; 
[8] http://www.jach.hawaii.edu/JCMT/surveys/Cosmology.html;
[9] \cite{Garn+08a}; 
[10] \cite{Grant+10}
[11] \cite{Feruglio+08};
[12] \cite{Berta+06,Berta+08};
[13] P.I. G. Covone \& M. Vaccari, http://people.na.infn.it/~covone/voice/voice.html
[14] Jarvis et al.\, (2012), in prep;
[15] \cite{Norris+06}, \cite{Middelberg+08}; 
[16] \cite{Wilkes+09}; 
[17] \cite{OM08}; 
[18] \cite{Garn+08b}; 
[19] \cite{Garn+10};
[20] \cite{Ibar+09};
[21] \cite{Lehmer+05}; 
[22]  http://www.astro.caltech.edu/$\sim$bsiana/cdfs\_opt; 
[23] Garching-Bochum Deep Survey, \cite{Hildebrandt+06}; 
[24] \cite{LeFevre+05}; 
[25] \cite{Damen+09}, irsa.ipac.caltech.edu/data/SPITZER/docs/spitzermission/observingprograms/legacy/; 
[26] Survey of the CDFS with the Large Apex Bolometer Camera (LABOCA), \cite{Weiss+09}; 
[27] \cite{Miller+08};
[28] \cite{Pierre+07}; 
[29] \cite{Ueda+08}; 
[30] www.cfht.hawaii.edu/Science/CFHTLS; 
[31] \cite{Iovino+05}; \cite{Temporin+08}; 
[32] \cite{Bondi+07};
[33] \cite{Simpson+06};
[34] \cite{Furusawa+08};

\vspace{0.5cm}
\end{table*}

\subsection{Optical surveys}
\label{subsec:opticalsurveys}

A number of optical surveys overlap one or more of the SERVS fields. Among
the most significant are the ESO/\emph{Spitzer} Imaging Survey (ESIS) in
ES1 \citep{Berta+06,Berta+08} which reaches depths of 25, 25, 24.5, 23.2
(Vega) in $B, V, R$ \& $I$, respectively, the Canada-France-Hawaii Telescope
Legacy Survey (CFHTLS; both deep and wide pointings in $u^{*}, g, r, i$ \&
$z$ in the XMM-LSS field reaching depths of 28.7, 28.9, 28.5, 28.4, 27.0 and
26.4, 26.6, 25.9, 25.5, 24.8 AB magnitudes, respectively), and ancillary
SWIRE data in Lockman, EN1 and CDFS in a variety of depths and filters,
but typically reaching at least $r=24.5$ \citep{Gonzalez+11}. 

The {\em Spitzer} Adaptation of the Red Sequence Cluster Survey
(SpARCS\footnote{http://www.faculty.ucr.edu/$\sim$gillianw/SpARCS},
\citealt{Muzzin+09}, \citealt{Wilson+09}, DeGroot et al., 2012, in prep,
Muzzin et al., 2012a, in prep) has imaged the entire SWIRE area (excluding
the XMM-LSS field which is covered by the CFHTLS) in the $z'$ filter using
MegaCam on the Canada France Hawaii Telescope (CHFT) or the Mosaic camera on
the Blanco Telescope at Cerro Tololo Inter-American Observatory (CTIO). The
MegaCam observations reach a mean depth of 24.2 AB magnitudes, and the Mosaic
camera observations reach a mean depth of 24.0 AB magnitudes. The SpARCS
collaboration has spectroscopically confirmed $\sim15$ clusters at $z\geq1$
(\citealt{Muzzin+09}, \citealt{Wilson+09}, \citealt{Demarco+10},
\citealt{Muzzin+11}, Muzzin et al., 2012b, in prep, Wilson et al., 2012a, in
prep), and 
is in the process of carrying out detailed multipassband and spectroscopic
follow-up studies (Rettura et al., 2012, in prep, Lidman et al., 2012, in
prep, Noble et al., 2012, in prep, Ellingson et al., 2012, in prep,  Wilson
et al., 2012b, in prep). 

Started in October 2011, the VST/VOICE survey 
(P.I. G. Covone \& M. Vaccari) is surveying the CDFS and ES1 fields in
$u, g, r, i$ aiming at reaching AB$ \sim 26$ at 5$\,\sigma$. 

Optical spectroscopy has thus far been confined to small regions of SERVS -
such as the recently completed PRIMUS (PRIsm MUlti-object Survey,
\citealt{Coil+11}) survey, covering parts of the ES1, CDFS \& XMM-LSS fields
- or to specific types of objects. The largest spectroscopic survey is the
VVDS, which has $\approx$\,10\,000  
spectroscopic redshifts for field galaxies with $17.5<I_{\rm AB}<24$ 
in their XMM-LSS and CDFS subfields \citep{LeFevre+05}. 
In addition, spectra of AGN (Active Galactic Nuclei) and quasars, now totaling several hundred objects
selected by various techniques,
have been obtained in the SERVS 
fields by \cite{Lacy+07} , \cite{Trichas+10} and Lacy et al. (2012, in prep).

\subsection{Ground-based near-infrared surveys}
\label{subsec:NIRground}

One of several major surveys to be carried out by the Visible and Infrared Survey Telescope
for Astronomy (VISTA) is the five filter 
near-infrared VISTA Deep Extragalactic Observations (VIDEO) survey
(Jarvis et al.\ 2012, in preparation), which 
will cover 12 deg$^2$ to AB magnitudes of 25.7, 24.6, 24.5, 24.0 and 23.5 in $Z,
Y, J, H$ and $K_s$ filters. 
The ES1, XMM-LSS and CDFS SERVS fields are
designed to exactly overlap their corresponding VIDEO fields. The combination of VIDEO and 
SERVS will be a particularly potent tool for the study of galaxy evolution at high redshifts.

The Deep eXtragalactic Survey (DXS) is part of the UKIRT Infrared Deep Sky Survey
(UKIDSS, \citealt{Lawrence+07}), and will cover the
Lockman and EN1 fields to 23.1 and 22.5 (AB) in $J$ and $K$ respectively. As
of October 2011, Data Release 7 is available to the entire astronomical 
community. It covers parts of the 
EN1, Lockman and XMM-LSS fields. Data Release 9 is available to the community served by the 
European Southern Observatory, and includes additional data.

\subsection{Mid- and far-infrared, and submillimeter surveys}

The SERVS fields were designed to be contained within the SWIRE fields. This has mostly been 
achieved, although
constraints from other surveys mean that a small fraction of SERVS lies outside of the SWIRE 
coverage. 

The HerMES\footnote{http://hermes.sussex.ac.uk/} survey is a Herschel Key Project to survey most of the
extragalactic {\em Spitzer} fields, including the SWIRE fields. 
HerMES has six levels, corresponding to increasing depths, level 6 being the
shallowest. \cite{Smith+12} give measured flux densities
at which 50\% of injected sources result in good detections at the
SPIRE wavelengths of $(250, 350, 500) \, \mu$m ranging from (11.6, 13.2, 13.1)
mJy to (25.7, 27.1, 35.8) mJy, depending on the depth of the observation,
with the deeper observations being confusion limited. 
All of SERVS is covered to Level 6 or deeper, with
significant areas as deep as Level 3 ($5\,\sigma$ limiting flux density
$\approx$\,7\,mJy at $160\, \mu$m with the PACS\footnote{Photodetector Array Camera \& Spectrometer} instrument).
Full details of the HerMES survey are given in \cite{Oliver+12}. 

The wide area component of the SCUBA-2 Cosmology Survey (S2CLS) 
will cover the XMM-LSS, Lockman, EN1 and CDFS fields. 
The survey will be performed at $850\, \mu$m to a root
mean square (rms) noise of 0.7 mJy.

\subsection{Radio surveys}

The Australia Telescope Large Area Survey (ATLAS; \citealt{Norris+06}) 
overlaps with much of the SERVS fields in ES1 and CDFS. The ATLAS survey will
have an rms of $10\, \mu$Jy and a spatial resolution of 
$\approx$\,8$''$ at $1.4\,$Ghz across both fields (Banfield et al. in prep).  
A preliminary release of the survey data for CDFS \citep{Norris+06} and ES1
\citep{Middelberg+08} has an rms noise of $30\, \mu$Jy. An image showing the
overlap between the SERVS and ATLAS fields can be found in the accompanying
paper by \cite{Norris+11}. Subsequent ATLAS data releases will be published
shortly by Hales et al. (2012) and Banfield et al. (2012), both in
preparation.  

The EN1 and Lockman fields have been surveyed at $610\,$MHz with the Giant Meter
Wave Telescope (GMRT; \citealt{Garn+08a,Garn+08b}, respectively).
These surveys have a mean rms of $\approx$\,60\,$\mu$Jy at a spatial resolution of $\approx$\,6$''$.

The Very Large Array (VLA) has conducted several surveys in the SERVS fields.
The Lockman field includes the deepest radio survey at $1.4\,$GHz to date, 
a single $40'\times 40'$ 
pointing centered at 161.5d, $+$59.017d, reaching a $5\,\sigma$ detection limit of $15\, \mu$Jy near 
the center of the primary beam (Owen \& Morrison 2008), and further deep coverage
by \cite{Ibar+09}. \cite{Simpson+06} have surveyed the Subaru Extragalactic Deep Survey region with the VLA
to a detection limit of $100\, \mu$Jy, and \cite{Bondi+07} have surveyed the
VVDS field in XMM-LSS at both 610MHz with the GMRT and $1.4\,$GHz with the VLA to limits of 
$\approx$\,200 and $80\, \mu$Jy, respectively. 
The Faint Images of the
Radio Sky at Twenty cm (FIRST, \citealt{WBHG97}) survey covers the Lockman,
EN1 surveys and part of the XMM-LSS survey 
at $1.4\,$GHz to a sensitivity limit of $\approx$\,1\,mJy at a spatial resolution of $\approx$\,5$''$.

Two pathfinder telescopes for the proposed Square Kilometer Array are currently under construction
in the southern hemisphere. Both of these telescopes will undertake continuum surveys that will
cover the southern SERVS fields. The Evolutionary Map of the Universe
(EMU, \citealt{NorrisEMU+11}) survey with the 
Australian Square Kilometer Array Pathfinder (ASKAP) will cover the whole southern sky to 10\,$\mu$Jy
rms sensitivity at $1.4\,$GHz with a $10''$ FWHM (full width at half-maximum)
synthesized beam. The South African 
Karoo Array Telescope (MeerKAT, \citealt{Jonas+09}) will conduct the 
MeerKAT International Giga-Hertz Tiered Extragalactic Exploration (MIGHTEE) survey, which has a
strong SERVS participation. MIGHTEE has several tiers, one of which will include the southern 
SERVS/VIDEO fields to $1\, \mu$Jy rms at $1.4\,$GHz with an $\approx$\,5$''$ FWHM beam. In the north,
the Low Frequency Array (LOFAR, \citealt{Rottgering+11}) will target the
SERVS/SWIRE fields for deep surveys at frequencies 
of $\sim 100\,$MHz. In the northern hemisphere, the WODAN (Westerbork Observations
of the Deep APERTIF northern-Sky) is being proposed for Westerbork+APERTIF
and will match EMU sensitivity and resolution \citep{Rottgering+11}.

\subsection{X-ray surveys}

The XMM-LSS field overlaps with the XMM-LSS survey \citep{Pierre+07} 
and the Subaru/XMM-Newton Deep Survey (SXDS, \citealt{Ueda+08}).
\cite{Wilkes+09} have a deep {\em Chandra} survey overlapping with the 
deep VLA pointing of \cite{OM08} in Lockman. In EN1, {\em Chandra} program
690062 (P.I.\ Nandra) covers $\approx$\,1\,deg$^2$ of the central portion of
the field.

\subsection{Further ground-based data taken or to be taken by the SERVS team}

\begin{figure}
\centering
\hspace{-0.5cm}
\includegraphics[scale=0.6]{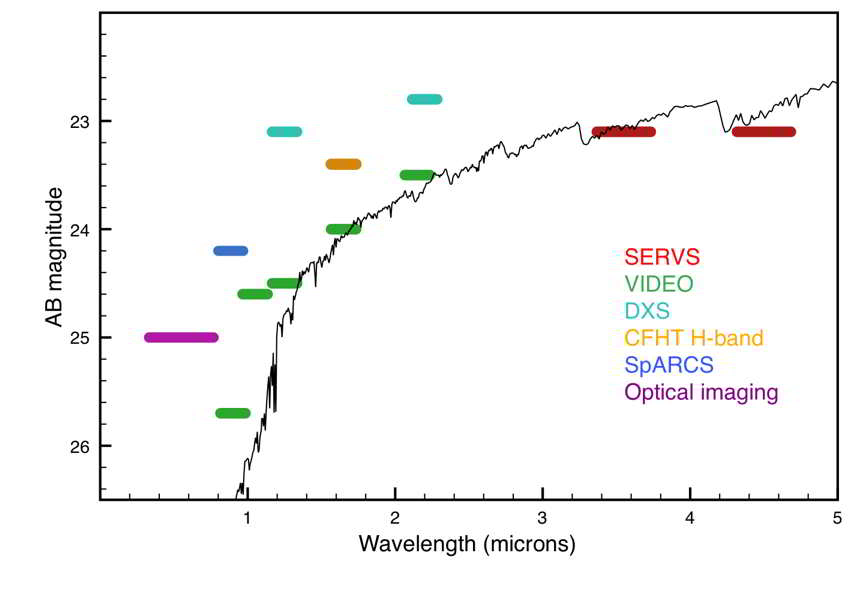}
\caption{Approximate $5\,\sigma$ depths of SERVS and near-IR and optical
  surveys covering the same areas. Overplotted is the SED of a 1Gyr old stellar population
  at $z=2$ from \cite{Maraston05}. SERVS is in \emph{red}, VIDEO in
  \emph{green}, DXS in \emph{cyan}, CFHT $H$-band in \emph{orange} and SpARCS
  in \emph{blue}. Our target depth for optical bands imaging is shown as the
  magenta bar (several fields already have imaging to at least this
  depth). \emph{Note that no ground-based survey covers the entirety of the SERVS data, hence
  multiple surveys are overlaid}.} 
\vspace{0.5cm}
\label{fig:depths}
\end{figure}

Further multiwavelength 
ancillary data on the SERVS fields are currently being obtained with 
the overall goal of matching the SERVS depth in shorter wavebands. 
These data will be made available as part of the overall SERVS public
release. Observations are concentrated on longer optical and near-IR
wavebands as these are generally more useful for photometric redshift
estimates at higher redshifts
\citep{vanDokkum+06,BvDC08,Ilbert+09,Cardamone+10}. In the optical, the SDSS
filter set is used where possible, as the narrower bands allow
higher-fidelity photometric redshifts than the Johnson-Cousins system (see 
Figure~\ref{fig:zphot_ex2}). 

Target depth in the 
optical is an AB magnitude $\approx$\,25 and $\approx$\,23 in the near-infrared
(the area covered by VIDEO will be significantly deeper than this). 
Figure~\ref{fig:depths} shows a comparison of the SERVS depth with those of the other
major optical/near-IR surveys planned or in progress.

Observations with SuprimeCam on the Subaru telescope have been carried out 
in $i$ and $z$ bands as part of the Gemini-Subaru
(PI A.\ Verma) and Keck-Subaru (P.I.\ S.A.\ Stanford) time swaps.
The $z$-band observations are concentrated in the northern fields, as VIDEO
will cover the southern fields in $Z$.

An optical imaging campaign of the ES1 and CDFS fields using the CTIO-4m
\emph{Mosaic} camera has been completed (in November 2009 \& October 2010,
with a total of 9 nights). The specific goals were (1) to complete imaging in
$r$ and $i$ to 24.2 and 23.2 (Vega) over the whole area of the ES1 and CDFS
fields and (2) to obtain deeper $r$- and $i$-band data (to 25.0 and 24.0) in
the center of ES1 to complement deep surveys by VISTA and warm \emph{Spitzer}, and
where deep data at shorter wavelengths exist from ESO. Images are currently
being processed. 

An $H$-band imaging campaign, led by M.\ Lehnert has obtained 
$H$-band data with WIRCAM on the 
CFHT, to match the UKIDSS DXS $J$ and $K$ imaging in the EN1 field. These data are
currently being analyzed.

\subsection{Photometric redshifts}
\label{subsec:photoz}

The wealth of ground-based data available in the SERVS fields in conjunction
with the IRAC data obtained through the SERVS project provides the ideal
basis for a robust determination of photometric 
redshifts, stellar masses and other stellar population properties
from SEDs. Here the photometric
redshifts were computed via SED fitting
using the HyperZ code \citep{Bolzonella+00} and the \cite{Maraston05}
stellar population templates, following the procedure outlined in Pforr,
Maraston \& Tonini (2012b, in prep). In brief, the best fit is determined
by minimizing the $\chi^2$ between a large grid of
population model templates (for various star formation histories, ages,
metallicities, reddening, and redshifts) and the photometric data.
Specifically, we use a template set with exponentially declining star
formation rates, metallicities ranging from 1/5 to twice solar, a
minimum age of 0.1 Gyr 
and a \cite{Calzetti+00} reddening law with
extinction parameter A$_V$ ranging between 0 and 3 mag. This template set
gives the minimum variance between photometric redshifts and the
spectroscopic redshifts calibration set. 

By comparing to available spectroscopic redshifts using the compilation of
redshifts in the SWIRE fields by Vaccari et al. (2012, in preparation), a
redshift accuracy of  $\Delta z/(1+z)=0.011\pm0.072$ is obtained for a
dataset with an optimal 
wavelength coverage (i.e. U, g, r, i, Z, J, K, IRAC1 [3.6], IRAC2 [4.5] for
the EN1 field). Due to the small\footnote{In the EN1 field, among the 2997 extended sources with optimal
wavelength coverage (respectively 8513 with nonoptimal wavelength coverage),
only 30 (respectively 60) have spectroscopic redshifts.} number of SERVS objects with available
spectroscopic redshifts, we further confirmed the redshift accuracy by 
cross-matching the VVDS survey in the extended Chandra Deep Field
\citep{LeFevre+05} with the IRAC SIMPLE survey \citep{Damen+09}, 
obtaining 831 matches with redshifts and 4.5\,$\mu$m  
fluxes $>3\mu$Jy. However, the diversity inherent 
to the ancillary data coverage for the SERVS fields impairs the redshift
determination particularly for objects with a narrower wavelength
coverage. This is explored in detail in Pforr et al. (2012b, in preparation).
In summary, photometric redshifts display larger scatter when the
wavelength coverage used for the fitting is narrow, i.e. does not incorporate
the rest-frame UV, optical or near-IR rest-frame or a combination thereof,
particularly when spectral breaks such as the 
4000\AA\,break and the Lyman break are not covered by the filter setup
(see \citealt{Ilbert+09} and \citealt{Bolzonella+00} for studies on
photometric redshift accuracies).

\begin{figure}
\resizebox{\hsize}{!}{\includegraphics{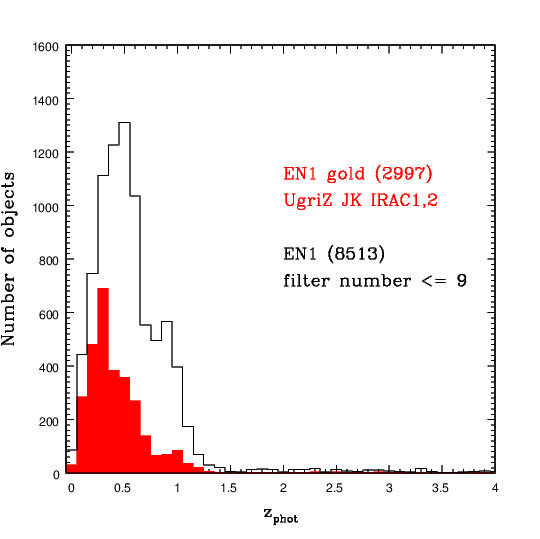}}
\caption{Photometric redshift distribution of SERVS galaxies in EN1 (\emph{black
  histogram}). Photometric redshifts are obtained from SED fitting using the
  HyperZ code and the \cite{Maraston05} stellar population synthesis
  templates. Included are only objects classified as extended in the optical
  and near-IR bands in order to exclude stars and AGN-dominated
  sources. Additionally the \emph{red histogram} shows the subsample of sources
  which are detected in all filter bands (U,g,r,i,Z,J,K, IRAC1, IRAC2 for
  EN1) which provides the most robust photometric redshift estimates. From
  Pforr, Maraston \& Tonini (2012b, in preparation).} 
\vspace{0.2cm}
\label{fig:zphot_ex2}
\end{figure}

The distribution of photometric redshifts obtained for extended\footnote{The
  decision whether an object is 
extended is based on stellarity flags provided with the ancillary data
in optical and near-IR filter bands. Sources with a pointlike nature 
indicative of stars or AGNs were excluded since the model template we use
in the fitting does not include nonthermal emission} objects, i.e. galaxies, in
EN1 is shown in Figure~\ref{fig:zphot_ex2}. The red histogram highlights the
most robust sample (so-called ``\emph{gold}'' sample) 
with the best wavelength coverage, i.e. corresponding to the case where the
object is detected in all filter bands. The black histogram includes
also objects that were detected in less photometric bands.

\vspace{0.5cm}
\section{Ongoing science with SERVS}
\label{sec:science_goals}

In this Section, we summarize already published or ongoing work by SERVS team
members to underline some of the key science goals of the SERVS survey. 
\vspace{0.2cm}
\subsection{Galaxies and their environments}
 
Sampling $\sim 0.8$\,Gpc$^3$, SERVS is
large enough to contain significant numbers of objects while still being
deep enough to find $L^*$ galaxies out to $z\approx 4$ (Figure~\ref{fig:zdist}).
By combining the five different fields of SERVS, the survey
effectively averages over large-scale structure, and is able to present a true picture
of the average properties of galaxies in the high redshift Universe.

Galaxy-galaxy correlations are being computed by van Kampen et al.\ (2012b,
in preparation)
in the SERVS fields. The five large, well-separated, SERVS fields 
enable us to average out the effects of large-scale structure on such 
measurements. Initial results on the EN1 field show the evolution of the
correlation function between high redshifts ($z > 1.3$) and intermediate
redshifts ($\sim 0.8$) using simple $[3.6]-[4.5]$ color cuts.

SERVS is deep enough and wide enough to find field clusters at $z>2$, should
they exist in significant numbers. The SERVS fields lie within the SpARCS fields
(see survey description in Section~\ref{subsec:opticalsurveys}, DeGroot et
al. 2012, in prep), The deep SERVS observations will allow both a more
accurate measurement of the faint end of the luminosity function of known
SpARCS clusters which fall within the SERVS footprint, and also the detection
of new clusters at higher redshifts than possible with SpARCS.  

In addition, Geach et al.\ (2012, in preparation) are pursuing a cluster selection
technique using photometric redshifts combined with Voronoi tessellation in
an attempt to identify further, mostly lower mass, cluster candidates.

\subsection{The Active Galactic Nuclei engine}

The [3.6] and [4.5] bands 
are important diagnostics of AGN SEDs, as they are where 
host galaxy light and hot dust emission from the torus 
overlap in the SEDs of many dust-obscured AGN and
quasars at moderate redshifts ($z\sim 1$). 
In unobscured, or lightly obscured 
objects, this is where the optical/UV emission from the accretion disk
transitions to the hot dust emission. \cite{Petric10} \& Petric et al. (2012,
in preparation) present SEDs of AGNs and quasars
selected in the mid-infrared, and use SERVS data to help apportion the
different sources of near-infrared light. The luminosities of the hosts
themselves, if free from contamination by AGN-related light, can also be used
to study the stellar masses of the host galaxies.

\subsection{AGN and their environments}

Current models for galaxy formation indicate that AGN and quasar activity
play an important role in galaxy formation (e.g. \citealt{Hopkins+06}),
regulating the growth of their host galaxies through feedback (see for
example \citealt{Schawinski+07}, or more recently \citealt{Farrah+12}, who
provide new and strong evidence for AGN feedback). However, the exact nature
of this feedback process is unclear. Environments in which AGN and quasars
lie can indicate the masses of the dark halos they inhabit, and also how
these masses depend on AGN luminosity and redshift \citep{Farrah+04}. These can illuminate
models for feedback, for example, a preponderance of AGN in massive halos,
accreting at relatively low rates might be an indicator that their host
galaxies are no longer growing rapidly \citep{Hopkins+07}. At low redshifts
($z<0.6$) the SDSS has been used to successfully perform these experiments
\citep{Padmanabhan+09}. SERVS is able to take these studies to $z>>1$.  

\begin{figure}
\resizebox{\hsize}{!}{\includegraphics[trim=0.5cm 0cm 0cm 0cm,clip=true]{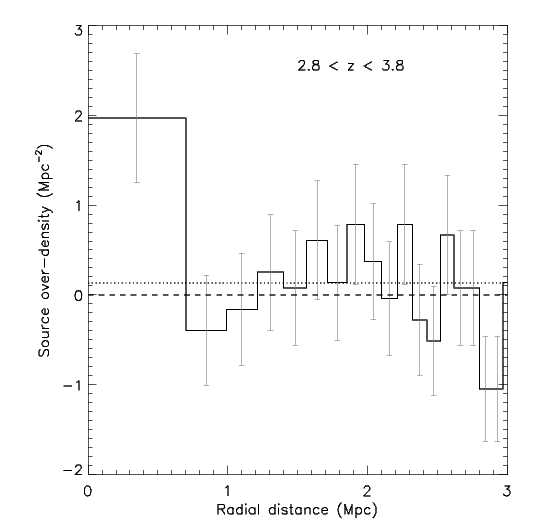}}
\caption{ Stacked source over-density vs radial distance for the 
11 QSOs in the redshift range of $2.8 < z < 3.8$. The first bin 
has a radius of 700 kpc and the other bins are of the same area 
as the first. The error bars show the Poisson error on the number 
counts. The dashed line shows the subtracted local background 
level (zero level) determined from an annulus of $2 {\rm Mpc}^2$ ($400''$) 
from the QSOs. The dotted line shows, for comparison, the global 
background as determined from taking the average source density 
in large apertures over the SERVS fields. This is the source density 
before being corrected for completeness. Figure from \cite{Falder+11}.}
\vspace{0.5cm}
\label{fig:falder_env}
\end{figure}

One particular area where SERVS is uniquely valuable is in determining the
environments of high redshift AGN.  \cite{Falder+11} find significant
($>4\,\sigma$) overdensity of galaxies around 
QSOs in a redshift bin centered on $z \sim 2.0$ and an ($>2\,\sigma$)
overdensity of galaxies around QSOs in a redshift bin centered on $z \sim
3.3$ (see Figure~\ref{fig:falder_env}). 

Nielsen et al. (2012, in preparation) are investigating the environments of AGN and
quasars selected in the mid-infrared. For the first time the environments of
luminous quasars at $0.8 < z \stackrel{<}{_{\sim}}3$ are being characterized,
enabling a comparison of the environments of dust obscured and normal quasars
at these redshifts. 

\subsection{High-z quasar searches}

The unique multiband
SERVS data set will be a valuable tool for constraining the faint end
of the quasar luminosity function at high redshifts. Quasar searches have
been or will be carried out in the SERVS fields using the combination of
SERVS, DXS, VIDEO and SpARCS data, in addition to the SERVS 
CTIO and Subaru ancillary data. This large range of wavelengths 
allows for the rejection of many
contaminants of high-$z$ quasar searches on the basis of near-infrared
photometry alone. Current estimates of the high-$z$ quasar luminosity
function (e.g. \citealt{Willott+10} suggests somewhere between
3-14 quasars at $5.5<z<6.5$ and 1-5 quasars at $6.5<z<7.5$. The large
range is due to the uncertainty in the faint-end slope of the quasar 
luminosity function, and SERVS will be able to constrain this well. 

\vspace{0.2cm}
\subsection{Infrared-faint radio sources}

\begin{figure}
\resizebox{\hsize}{!}{\includegraphics{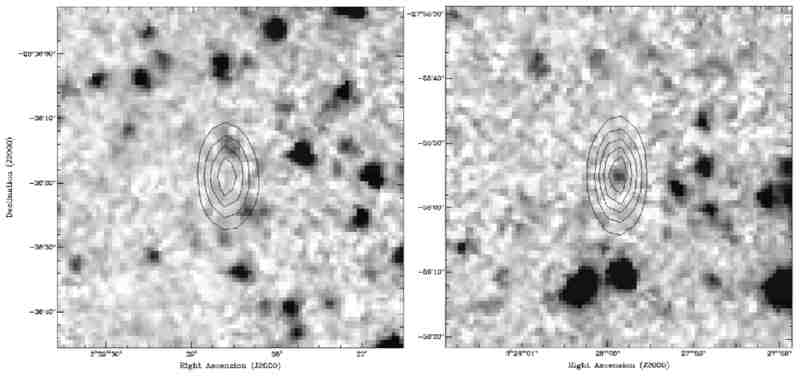}}
\caption{Two representative infrared-faint radio sources (IFRS). The greyscale is the [3.6] SERVS data, and 
the contours are the 20 cm image, with contour levels of (1, 2, 3, 4, 5) mJy/beam. The left 
hand image is a non-detection(CS0194) and the right-hand image is a candidate detection 
(CS0114), Figure 2. of \cite{Norris+11}.}
\vspace{0.5cm}
\label{fig:norris_ifrs}
\end{figure}

\cite{Norris+06} describe the discovery of infrared-faint radio sources
(IFRS), a population of radio sources with
host galaxy fluxes well below the limit of the SWIRE survey. \cite{Huynh+10} 
used deep IRAC data to place even more stringent limits on them, and
concluded that these are most likely radio-loud AGN with faint host galaxies,
but the sample to date is small. \cite{Norris+11} present an initial study of
this hitherto unsuspected population with SERVS, including stacking of
objects that are too faint to be detected, even in SERVS, and that may
represent very high redshift radio-loud galaxies, possibly suffering from
significant dust extinction (see Figure~\ref{fig:norris_ifrs}).

Using a combination of SERVS data and GMRT/VLA radio observations of the
Lockman Hole at 610 MHz and 1.4 GHz, \cite{Afonso+11} study a sample of Ultra
Steep Spectrum (USS) radio sources and suggest the likely existence of higher
redshifts among the submillijansky USS population, raising the possibility that the
high efficiency of the USS technique for the selection of high-redshift
sources remains even at the sub-mJy level.  

\vspace{0.5cm}

\subsection{Obscured star formation}

Although SERVS cannot be used as a direct indicator of obscured star
formation on its own, overlap with surveys
by SCUBA-2 (S2CLS) and {\em Herschel} (HerMES) will allow more reliable source identification than 
possible using shorter wavelength data, and better characterization of any extinction of the
stellar light, as well as the stellar mass of the galaxy. Based on recent simulations
of {\em Herschel} observations, we expect to detect $\sim 700$ unconfused sources per 
deg$^2$ in $\geq 3$ {\em Herschel} bands \citep{FLPD08}.
At the SERVS depths, we expect to detect $>95$\% of
these sources in both IRAC bands, and with the aid of our ancillary optical and 
near-infrared data, obtain photometric redshifts and stellar masses for $\sim
12000$ sources. This will be sufficient to study trends in star formation
rate with stellar mass and redshift, for example, to test the idea of
``downsizing'' of the most actively star-forming galaxies.  

\vspace{0.2cm}

\section{Summary}
\label{sec:summary}

The \emph{Spitzer} Extragalactic Representative Survey (SERVS) is designed to
open up a medium-depth, medium-area part of parameter space in the
near-infrared, covering $18\,\rm{deg}^2$ to $\approx$\,2$\,\mu$Jy in the {\em Spitzer}
[3.6] and [4.5] bands in five highly observed astronomical fields (EN1,
ES1, Lockman Hole, CDFS and XMM-LSS). The five SERVS fields are centered on
or close to those of corresponding fields surveyed by the shallower SWIRE
fields, and they overlap with several other major surveys covering wavelengths
from the X-ray to the radio. Of particular importance are 
near-infrared data, as these allow accurate photometric redshifts to be
obtained. SERVS overlaps exactly with the 12 deg$^2$ of the VISTA VIDEO survey
in the south, and is covered by the UKIDSS DXS survey in the north. SERVS
also has good overlap with HerMES in the far-infrared, which covers SWIRE and other {\em
  Spitzer} survey fields, with deeper subfields within many of the SERVS
fields. Sampling $\sim 0.8$\,Gpc$^3$ and redshifts from 1 to 5, the survey is
large enough to contain significant numbers of rare objects, such as luminous
quasars, ultra-luminous infrared galaxies, radio galaxies and galaxy
clusters, while still being deep enough to find $L^*$ galaxies out to
$z\approx 4$. In this article, we have described the \emph{Spitzer}
observations, the data processing and the wealth of ancillary data
available in the fields covered by SERVS. Mosaics and catalogs will be made
available to the community in the summer of 2012  through the {\em Infrared
  Science Archive} (IRSA).

\vspace{0.2cm}
\section{Acknowledgements}

This work is based on observations made with the {\em Spitzer Space
  Telescope}, which is operated by the Jet Propulsion Laboratory (JPL), California
Institute of Technology (Caltech), under a contract with NASA. Support for this work
was provided by NASA through an award issued by JPL/Caltech. The National
Radio Astronomy Observatory is a facility of the National Science Foundation
operated under cooperative agreement by Associated Universities, Inc. J.A.,
H.M., M.G. and 
L.B. gratefully acknowledge support from the Science and Technology Foundation
(FCT, Portugal) through the research grant PTDC/FIS/100170/2008 and the
Fellowships SFRH/BD/31338/2006 (HM) and SFRH/BPD/62966/2009
(L.B.). G.W. gratefully acknowledges support from NSF grant AST-0909198. This
publication 
makes use of data products from the Two Micron All Sky Survey, which is a
joint project of the University of Massachusetts and the Infrared Processing
and Analysis Center/California Institute of Technology, funded by the
National Aeronautics and Space Administration and the National Science
Foundation.

\setcounter{figure}{0} \renewcommand{\thefigure}{A.\arabic{figure}} 
\setcounter{table}{0} \renewcommand{\thetable}{A.\arabic{table}} 
\vspace{1cm}
\appendix
\vspace{0.5cm}
\section{Number counts for the EN1, ES1, Lockman \& CDFS fields}
\label{app:fourfieldsnumcounts}

Number counts were derived for the five fields. Source colors were also
compared to 2MASS (see details in Section~\ref{subsec:data_analysis}). We
present here the four remaining fields: EN1, ES1, Lockman and CDFS. All of
them show consistent number count features and completeness levels. SERVS bright
stars colors for all of the fields are coherent with 2MASS. 
\vspace{1cm}

\begin{figure}[ht]
\begin{minipage}[b]{0.5\linewidth}
\centering
\begin{tabular}{cc}
\hspace{0.3cm}
\includegraphics[scale=0.35]{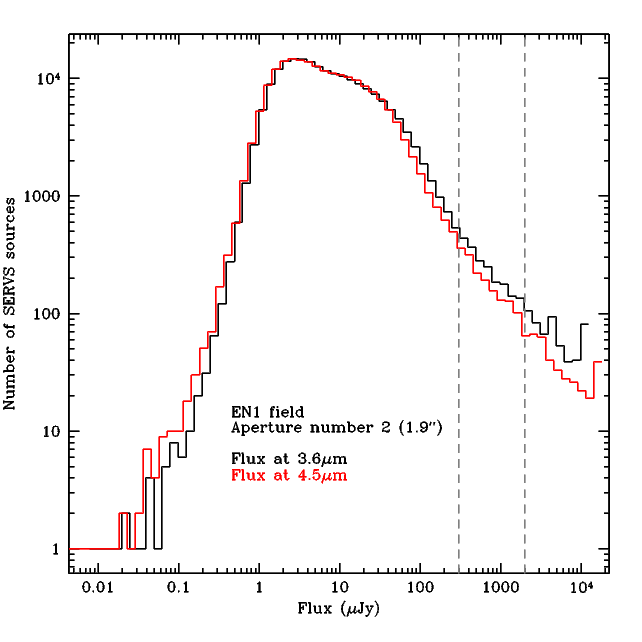} &
\hspace{0.2cm}
\includegraphics[scale=0.35]{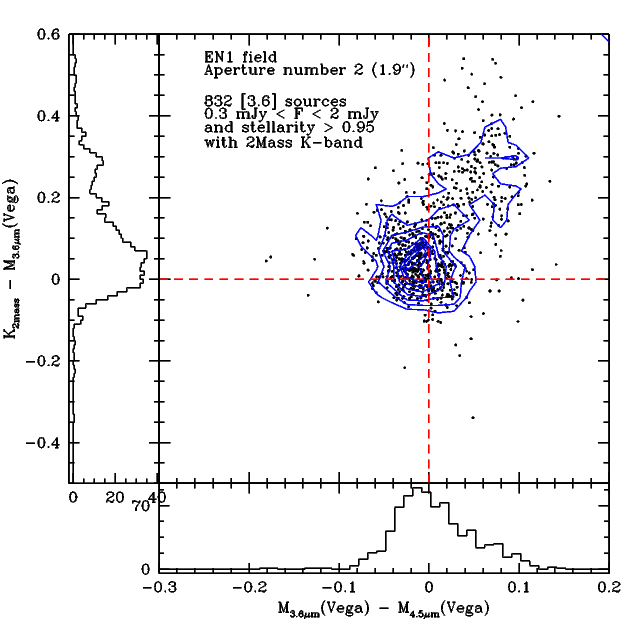}\\
\hspace{0.3cm}
\includegraphics[scale=0.35]{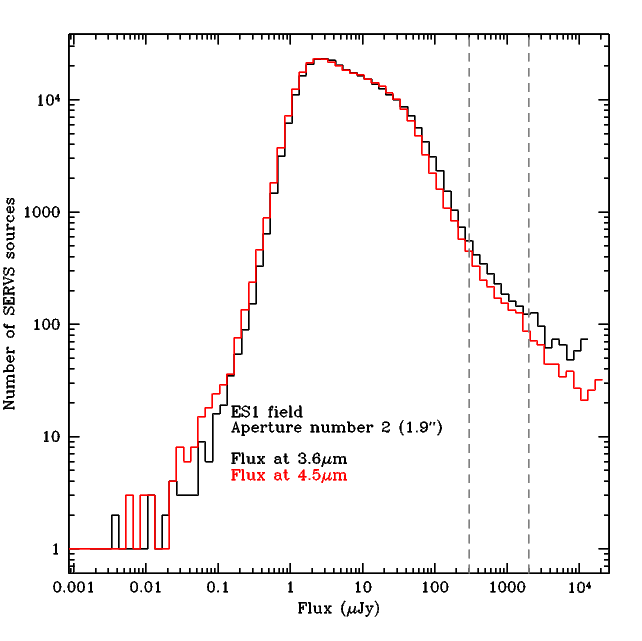} &
\hspace{0.2cm}
\includegraphics[scale=0.35]{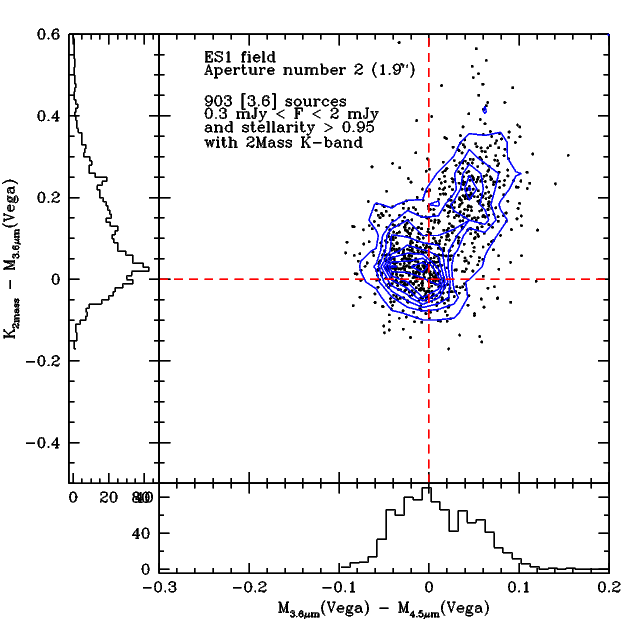}\\
\end{tabular}
\end{minipage}
\caption{\emph{Left side}: SERVS number counts versus flux at [3.6] and [4.5]
  for the EN1 (\emph{top}) and ES1 (\emph{bottom}) fields. Grey dashed lines
  show the selection of sources used in the bottom plot (with fluxes as
  $0.3\, < f < 2\,$mJy). \emph{Right side}: Color-color plot showing
  $K_{\rm{2MASS}} - M_{3.6\mu m}$(Vega) versus $M_{3.6\mu m}{\rm
    (Vega)}-M_{4.5\mu m}$(Vega) for sources within the flux range defined by
  the grey dashed lines above, plus a cut in stellarity index $>0.95$ and
  the existence of a 2MASS K-band measurement as an additional
  constraint. \emph{Red dashed} lines help pinpoint the location of the (0,0)
  point in this diagram.}
\vspace{0.2cm}
\label{fig:en1es1numcounts}
\end{figure}

\begin{figure}[ht]
\begin{minipage}[b]{0.5\linewidth}
\centering
\begin{tabular}{cc}
\hspace{0.3cm}
\includegraphics[scale=0.35]{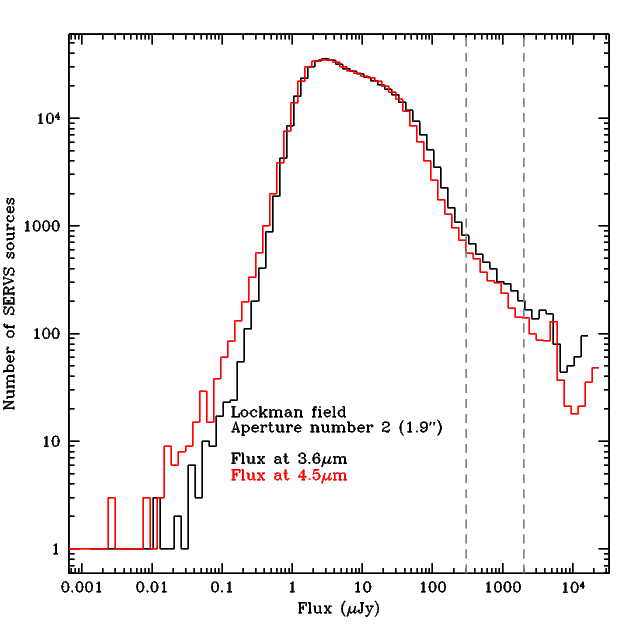} &
\hspace{0.2cm}
\includegraphics[scale=0.35]{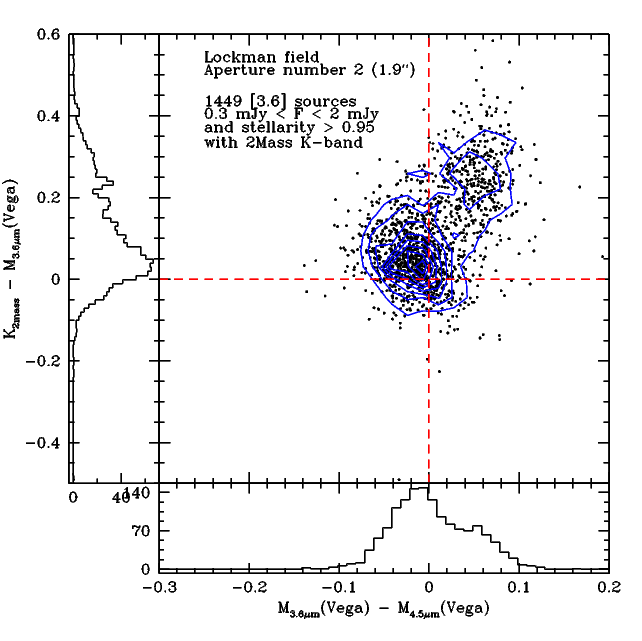}\\
\hspace{0.3cm}
\includegraphics[scale=0.35]{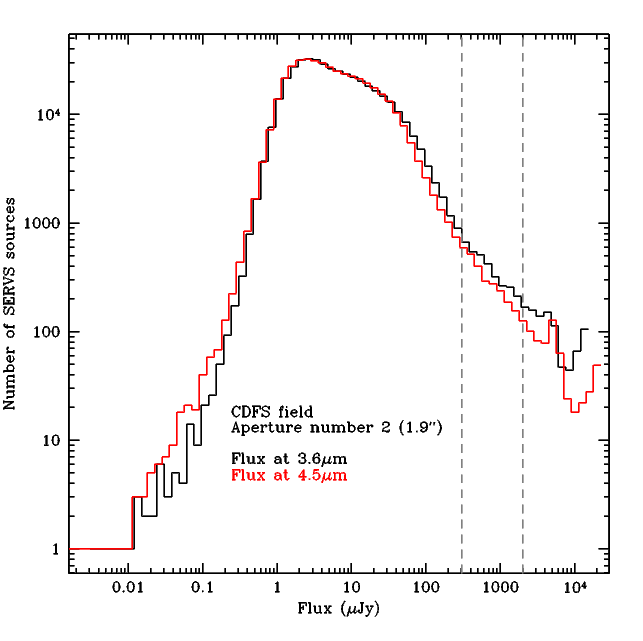} &
\hspace{0.2cm}
\includegraphics[scale=0.35]{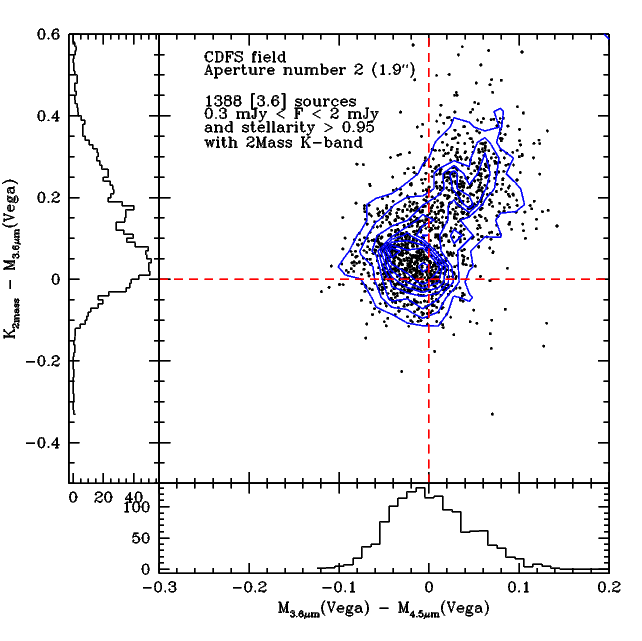}\\
\end{tabular}
\end{minipage}
\caption{\emph{Left side}: SERVS number counts versus flux at [3.6] and [4.5]
  for the Lockman (\emph{top}) and CDFS (\emph{bottom}) fields. Grey dashed lines
  show the selection of sources used in the bottom plot (with fluxes as
  $0.3\, < f < 2\,$mJy). \emph{Right side}: Color-color plot showing
  $K_{\rm{2MASS}} - M_{3.6\mu m}$(Vega) versus $M_{3.6\mu m}{\rm
    (Vega)}-M_{4.5\mu m}$(Vega) for sources within the flux range defined by
  the grey dashed lines above, plus a cut in stellarity index $>0.95$ and
  the existence of a 2MASS K-band measurement as an additional
  constraint. \emph{Red dashed} lines help pinpoint the location of the (0,0)
  point in this diagram.} 
\vspace{0.3cm}
\label{fig:lckmncdfsnumcounts}
\end{figure}

\vspace{1cm}

\section{{\sc SExtractor} \& {\sc Mopex} parameters used for the image
  processing and the catalogs extraction}
\label{app:sextractor}

{\sc Mopex} was used to process and co-add the raw images and {\sc SExtractor}
to extract information on the sources within.
In Table~\ref{table:sextractor} we list the core parameters used in both.
\vspace{2cm}

\begin{table}
\begin{center}
\vspace{0.1cm}
\caption{Values of the more important {\sc Mopex} \& {\sc SExtractor} parameters}
\begin{tabular}{llc}
\hline
\hline
Program (module)& Parameter & Value \\
\hline
{\sc Mopex} (DETECT)      &Detection\_Max\_Area& 100\\
{\sc Mopex} (DETECT)      &Detection\_Min\_Area& 0\\
{\sc Mopex} (DETECT)      &Detection\_Threshold& 4\\
{\sc Mopex} (MOSAICINT) & INTERP\_METHOD & 1 \\
{\sc Mopex} (MOSAICDUALOUTLIER)& MIN\_OUTL\_IMAGE &2\\
{\sc Mopex} (MOSAICDUALOUTLIER)& MIN\_OUTL\_FRAC &0.51\\
{\sc Mopex} (MOSAICOUTLIER) &THRESH\_OPTION& 1\\
{\sc Mopex} (MOSAICOUTLIER) &BOTTOM\_THRESHOLD& 0\\
{\sc Mopex} (MOSAICOUTLIER) &TOP\_THRESHOLD& 0\\
{\sc Mopex} (MOSAICOUTLIER) &MIN\_PIX\_NUM& 3\\
{\sc Mopex} (MOSAICRMASK)   &MIN\_COVERAGE&4\\
{\sc Mopex} (MOSAICRMASK)   &MAX\_COVERAGE&100\\
{\sc SExtractor} & DETECT\_MINAREA   & 5.0 \\
{\sc SExtractor} & DETECT\_THRESH    & 0.4\\
{\sc SExtractor} & ANALYSIS\_THRESH   & 0.4\\
{\sc SExtractor} & FILTER            & Default\\
{\sc SExtractor} & DEBLEND\_NTHRESH  & 64\\
{\sc SExtractor} & DEBLEND\_MINCONT  & 0.005\\
{\sc SExtractor} & SEEING\_FWHM      & 2.0\\
{\sc SExtractor} & BACK\_SIZE        & 32\\
{\sc SExtractor} & BACK\_FILTERSIZE  & 5\\
{\sc SExtractor} & BACKPHOTO\_TYPE   & LOCAL\\
{\sc SExtractor} & WEIGHT\_TYPE      & MAP\_WEIGHT\,$^*$\\
\hline\\
\end{tabular}
\label{table:sextractor}
\\\noindent {\it Notes}: Similar values can be found in \cite{Lacy+05} and
\cite{Ashby+09}\\
\noindent \hspace{-3.5cm} $^*$ Using coverage map as inverse variance weight map
\vspace{0.5cm}
\end{center}
\end{table}

\vspace{20cm}
\bibliographystyle{emulateapj}
\bibliography{servs}

\end{document}